\documentclass[reprint,superscriptaddress]{revtex4-2}

\usepackage{amsmath}
\usepackage{amssymb}
\usepackage{booktabs}
\usepackage{array}
\usepackage{multirow}
\usepackage{dcolumn}
\usepackage{bm}
\usepackage{graphicx}
\usepackage[caption=false]{subfig}
\usepackage[table]{xcolor}
\usepackage[colorlinks=true, linkcolor=blue, citecolor=blue, urlcolor=blue]{hyperref}
\usepackage{orcidlink}
\usepackage{siunitx}
\usepackage[section]{placeins}

\newcolumntype{B}{>{\bfseries}c}

\begin{document}

\title{Dirac oscillator in a helically twisted spacetime with axial torsion}

\author{Matheus D. Moro\orcidlink{0009-0009-8709-2414}}
\email[Matheus D. Moro - ]{matheusdinizmr01@gmail.com}
\affiliation{QPQI Group, Universidade Estadual de Ponta Grossa, 84030-900 Ponta Grossa, PR, Brasil}

\author{Fabiano M. Andrade\orcidlink{0000-0001-5383-6168}}
\email[Fabiano M. Andrade - ]{fmandrade@uepg.br}
\affiliation{QPQI Group, Universidade Estadual de Ponta Grossa, 84030-900 Ponta Grossa, PR, Brasil}
\affiliation{Departamento de Matem\'{a}tica e Estat\'{i}stica, Universidade Estadual de Ponta Grossa, \\84030-900 Ponta Grossa, PR, Brasil}

\author{Faizuddin Ahmed\orcidlink{0000-0003-2196-9622}}
\email[Faizuddin Ahmed - ]{faizuddinahmed15@gmail.com}
\affiliation{Department of Physics, The Assam Royal Global University, Guwahati, 781035, Assam, India}

\author{Edilberto O. Silva\orcidlink{0000-0002-0297-5747}}
\email[Edilberto O. Silva - ]{edilberto.silva@ufma.br}
\affiliation{Programa de P\'os-Gradua\c c\~ao em F\'{\i}sica \& Coordena\c c\~ao do Curso de F\'{\i}sica -- Bacharelado, Universidade Federal do Maranh\~{a}o, 65085-580 S\~{a}o Lu\'{\i}s, Maranh\~{a}o, Brazil}

\date{\today}

\begin{abstract}
We investigate the Dirac oscillator in a helically twisted spacetime endowed with a uniform axial torsion. Starting from an orthonormal coframe, we compute the Levi--Civita spin connection explicitly and separate the geometric contribution from the axial contortion. Retaining the matrix $\beta$ in the radial Moshinsky coupling, we show that the second-order problem is the ordered product $\hat\Pi_+\hat\Pi_-$ rather than the square of a single operator. The resulting radial dynamics is a coupled, self-adjoint two-component system in which the spin connection supplies the correct cylindrical radial operator, while the off-diagonal metric generates the helical combination $m/r-\omega k$ and a Coulomb-like geometric term. A finite-element solution reproduces the planar Dirac-oscillator spectrum in the flat limit and reveals asymmetric dependence on the longitudinal momentum, avoided level crossings, and a supersymmetric zero mode at $E=Mc^2$. The axial torsion and longitudinal momentum preserve this zero mode, whereas the helical twist lifts it quadratically. Sector-resolved thermodynamic functions are obtained from the relativistic bound-state spectrum. The explicit spinors further determine longitudinal vector and axial currents, and a Witten-index analysis identifies the helical twist as the deformation that removes the protected zero mode.
\end{abstract}

\maketitle

\section{Introduction}
\label{sec:intro}

Relativistic oscillator-type couplings in the Dirac equation go back to the early work of Ito, Mori, and Carriere~\cite{NCA.1967.51.1119}, while the model now known as the Dirac oscillator (DO) was formulated systematically by Moshinsky and Szczepaniak~\cite{JPA.1989.22.17.L817}. In the conventional approach to the Klein--Gordon oscillator, a harmonic confinement is introduced through the minimal-coupling substitution $\boldsymbol{p}\to\boldsymbol{p}-iM\omega_K\boldsymbol{r}$, where $\omega_K$ denotes the Klein--Gordon oscillator frequency. The Dirac oscillator adopts a different relativistic prescription, $\boldsymbol{p}\to\boldsymbol{p}-iM\omega_D\beta\boldsymbol{r}$, with $\beta=\gamma^0$ and $\omega_D$ the Dirac-oscillator frequency. The matrix $\beta$ is part of the standard Moshinsky coupling; in the present work, the important point is to keep this matrix structure explicitly when the curved-space Dirac equation is separated into its upper and lower Pauli components. This coupling leads to exact solvability in the flat model, hidden supersymmetry, two-body extensions, and spin-orbit effects without external magnetic fields~\cite{PRL.1990.64.14.1643,IJMPA.1991.6.9.1567,JPA.1991.24.3.667,JPA.1995.28.22.6447,IJMPA.2004.19.2765}. The same algebraic structure has motivated analyses of wave-packet dynamics and Foldy--Wouthuysen representations~\cite{JPA.1997.30.7.2585}, as well as the relation between the Dirac oscillator and Jaynes--Cummings-type quantum-optical models~\cite{JPA.1999.32.28.5367,PRA.2007.76.041801,JPA.2010.43.28.285204}. Its photonic and microwave realizations further established the DO as a bridge between relativistic quantum mechanics and experimentally accessible analogue systems~\cite{OL.2010.35.8.1302,PRL.2013.111.170405,JLTP.2022.209.1.44,Nature.2010.463.68}. 

Over the past three decades, the Dirac oscillator has proven to be a remarkably versatile theoretical laboratory for investigating fundamental aspects of relativistic quantum mechanics in various contexts. It has been studied in the presence of external electromagnetic and Aharonov--Bohm fields~\cite{PRA.2011.84.3.032109,PLA.2004.325.1.21}, non-commutative geometries~\cite{IJMPA.2011.26.29.4991,PLA.2012.376.36.2467}, minimal-length and deformed-algebra scenarios~\cite{JPA.2005.38.8.1747,PLB.2014.731.327,PLB.2014.738.44}, rainbow-gravity scenarios~\cite{EPJP.2018.133.10.409}, and topological defects~\cite{PRD.2020.102.10.105020,PS.2025.100.9.095308,PRA.2011.84.3.032109,AP.2015.355.48,AHEP.2017.2017.1.1723567}. The DO has also found applications in condensed-matter and analog platforms, particularly in graphene-like systems, graphene quantum dots, and Dirac/Weyl materials with strain- or geometry-induced gauge structures~\cite{RMP.2009.81.109,Science.2010.329.544,PLA.2016.380.773,PS.2015.90.4.045702,PRL.2011.107.12.127205,RMP.2018.90.015001}.

A particularly fruitful direction of research concerns the behavior of the Dirac oscillator in curved spacetimes. The interplay between geometry and quantum mechanical confinement reveals subtle physical effects that cannot be anticipated from flat-space intuition. Early investigations focused on maximally symmetric spaces such as de Sitter~\cite{JMP.2015.56.1.012101}, where the spherical symmetry allows for complete analytical solutions. Subsequently, attention has shifted toward less symmetric geometries that model realistic physical situations such as cosmic strings~\cite{JPA.2020.53.18.185204,EPJP.2012.127.7.82,EPJC.2019.79.311}, rotating frames~\cite{GRG.2013.45.1847}, Som--Raychaudhuri spacetime~\cite{GRG.2018.50.5.47}, and spacetimes with torsion within the framework of Einstein--Cartan gravity~\cite{NPB.2026.1029.117506}. In particular, Dirac-oscillator realizations have been analyzed in spinning and magnetic cosmic string backgrounds, in vector and scalar backgrounds with spin and pseudospin symmetries, in self-adjoint extension treatments of singular Hamiltonians, in $\kappa$-deformed kinematics and in related Lorentz-violating scenarios, including the models discussed in Refs.~\cite{EPJC.2014.74.3187,EPL.2014.108.30003,EPJC.2019.79.596,PRD.2020.102.10.105020,FrontPhys.2019.7.175,Universe.2020.6.11.203}. 

Among the curved backgrounds of physical interest, helically twisted spacetimes occupy a distinguished position. The simplest representative of this class is described by the line element
\begin{equation}
ds^2=-c^2dt^2+dr^2+r^2d\phi^2+(dz+\omega r\,d\phi)^2,
\label{eq:helix-metric-intro}
\end{equation}
where $\omega$ is a real dimensionless parameter that parametrizes the helical twist. This convention is natural for the form $dz+\omega r\,d\phi$, since $z$ and $r$ carry dimensions of length while $d\phi$ is dimensionless. The same helically twisted geometry has recently been analyzed in terms of geodesic dynamics, wave optics, and effective geometric torsion~\cite{PLB.2025.870.139944}. This geometry arises naturally in several physically relevant contexts. Spacetimes associated with screw-dislocation-type defects or disclination defects in condensed matter analogs exhibit helical structure~\cite{AP.1992.216.1.1,IJTP.1987.26.8.715}. In effective geometric descriptions of crystalline, liquid-crystal, photonic, and metamaterial systems, such terms model angular--axial mixing and torsion-like behavior induced by the medium. This geometric twist should not be confused with the independent axial torsion field introduced below in the spin connection. Recent developments in condensed matter systems such as topological materials, Dirac/Weyl semimetals, strain-engineered media, cold atoms, and photonic metamaterials allow for the simulation of effective curved spacetimes, including helical and torsion-like configurations~\cite{LRR.2011.14.3,NPB.2010.828.3.625,PRL.2011.107.7.075502,NatPhoton.2013.7.2.153,RMP.2008.80.885,Nature.2011.471.83}.

From a geometric standpoint, the metric~\eqref{eq:helix-metric-intro} is stationary and possesses cylindrical symmetry through invariance under time and longitudinal translations as well as under rotations $\phi\to\phi+\text{const.}$ Therefore the quantum numbers associated with $\partial_\phi$ and $\partial_z$, denoted by $m$ and $k$, remain well-defined separation constants. However, the metric is not diagonal in the standard cylindrical coordinate basis. The off-diagonal component $g_{\phi z}=\omega r$ encodes the helical mixing between angular and longitudinal displacements, so the radial dynamics depends on the angular--longitudinal combination $m/r-\omega k$ rather than on $m/r$ and $k$ separately. This coupling is the main geometric fingerprint of the helical background.

For fermions propagating in such a background, the curved-space Dirac equation involves the spin connection, also known as the Ricci rotation coefficients, which encodes how spinors are parallel-transported in the curved geometry. In the standard torsion-free Levi--Civita formulation, the spin connection is completely determined by the metric. However, in Einstein--Cartan theories or in effective models with intrinsic torsion~\cite{AP.1961.12.2.200,ARXIV.2006.gr-qc.0606062}, an additional contortion tensor arises, leading to extra spin-dependent interactions. One particularly interesting scenario is the inclusion of a purely axial torsion, which couples to the fermion's axial current $\bar{\psi}\gamma^\mu\gamma^5\psi$ and can mimic chiral anomalies or topological effects in condensed matter analogs~\cite{PRD.2018.98.025016,WS.1989.0356}.

Despite the growing interest in Dirac oscillators in curved and twisted spacetimes, several technical and conceptual issues remain inadequately addressed in the literature. Most treatments rely on symbolic manipulations or computational algebra systems to handle the spin connection, but the explicit matrix form of $\Omega_\mu^{(\mathrm{LC})}$ in the Dirac basis, decomposed into $2\times 2$ Pauli blocks, is rarely presented in full detail. This obscures the physical origin of various terms in the effective Hamiltonian. In the standard Dirac oscillator, the Moshinsky coupling always contains the Dirac matrix $\beta$. In curvilinear coordinates, the point is that this matrix structure must be retained throughout the decoupling, in particular in the radial replacement $p_r\to p_r-iM\omega_D\beta r$. After the spinor is split into upper and lower Pauli components, the $\beta$ matrix produces two different radial operators, $A_+=p_D+iM\omega_D r$ and $A_-=p_D-iM\omega_D r$. Consequently, the second-order problem involves the ordered product $\hat\Pi_+\hat\Pi_-$ rather than the square of a single operator. The cancellation referred to below is the cancellation between the cylindrical spin-connection contribution contained in $D_r=d/dr+1/(2r)$ and the oscillator ordering in $A_+A_-$; without this structure, one obtains incorrect radial derivative terms after decoupling. Furthermore, the relative contributions of purely geometric effects encoded in the Levi--Civita connection versus torsional or contortion effects have not been systematically disentangled. In particular, the Coulomb-like term proportional to $\omega k m/r$ that appears in the radial equation is often ambiguously attributed to curvature or torsion, whereas it arises from the non-diagonal metric through the anticommutator $\{\gamma^\phi,\gamma^z\}\neq 0$. Finally, a robust check of any curved-space formulation is the recovery of known flat-space results when the geometric deformation is removed. For the helical geometry, setting $\omega\to 0$ should yield the standard Dirac oscillator in flat cylindrical coordinates, without residual geometric artifacts.

In contrast with previous Dirac-oscillator studies in defect backgrounds, here the fermion is subject simultaneously to two independent geometric ingredients: the helical twist encoded in the off-diagonal metric component $g_{\phi z}=\omega r$, and the axial torsion introduced through the contortion sector of the spin connection. The former produces the angular--longitudinal mixing $m/r-\omega k$ and the associated Coulomb-like geometric term, whereas the latter enters as a constant axial spinorial shift. This separation allows us to identify which spectral features originate from the metric twist and which originate from genuine torsion.

The primary objective of this paper is to provide a complete, self-contained, and transparent derivation of the Dirac oscillator in the helically twisted spacetime~\eqref{eq:helix-metric-intro}. We focus on the explicit computation of the torsion-free Levi--Civita spin connection $\Omega_\mu^{(\mathrm{LC})}$ as $4\times 4$ matrices in the Dirac basis, along with their decomposition into $2\times 2$ Pauli blocks suitable for two-component spinor analysis. We carefully separate the geometric Levi--Civita contribution from the torsional contortion contribution, introducing spinorial shift operators $\hat{\mathcal{S}}_{\mathrm{LC}}(r)$ and $\hat{\mathcal{S}}_{\mathrm{ax}}$ that act explicitly in the first-order Dirac system. Through detailed operator algebra, we demonstrate the decoupling of the radial equation with step-by-step verification of key cancellations, particularly the absence of spurious $r\,\partial_r$ terms arising from the oscillator coupling. We identify the geometry-induced Coulomb-like contribution proportional to $\omega k m/r$, which originates purely from the non-vanishing anticommutator $\{\gamma^\phi,\gamma^z\}=+2\omega/r\,\mathbb{I}_4$ dictated by the metric. Finally, we verify the flat limit when $\omega\to 0$, confirming that the formalism reduces exactly to the known relativistic Dirac-oscillator result without residual artifacts.

The remainder of this paper is organized as follows. Section~\ref{sec:geometry} reviews the geometric setup, including the helical metric, the choice of orthonormal coframe, and the construction of curved gamma matrices. We compute the Levi--Civita spin connection using Cartan's structure equations and present the results both as $4\times 4$ matrices and in $2\times 2$ Pauli block form. Section~\ref{sec:torsion} introduces the purely axial torsion and the associated spinorial coupling. In Sec.~\ref{sec:dirac-oscillator}, we formulate the Dirac oscillator with Moshinsky coupling in the helical background, perform a separation of variables, and write the first-order two-component system. Section~\ref{sec:decoupling} derives the ordered second-order radial matrix equation and the explicit coupled radial system. Section~\ref{sec:numerical} presents the finite-element solution, its flat-limit validation, and the resulting spectra and densities. Section~\ref{sec:susy} discusses the supersymmetric zero mode and its response to the helical and torsional deformations. Section~\ref{sec:thermo} uses the computed spectrum to obtain sector-resolved thermodynamic functions. Section~\ref{sec:currents} evaluates the longitudinal vector and axial currents associated with the bound states. Section~\ref{sec:index} formulates the zero-mode protection in terms of the Witten index. Section~\ref{sec:nrlimit} derives the nonrelativistic limit and identifies the effective geometric and torsional terms that survive in this regime. Section~\ref{sec:conclusions} summarizes our conclusions.

\section{Geometric setup and Levi--Civita connection}
\label{sec:geometry}

We work with the helically twisted spacetime described by the line element
\begin{equation}
ds^2=-c^2dt^2+dr^2+r^2d\phi^2+(dz+\omega r\,d\phi)^2,
\label{eq:metric}
\end{equation}
where $(t,r,\phi,z)$ are the standard cylindrical coordinates and $\omega$ is a dimensionless constant that characterizes the helical twist. The metric tensor in matrix form reads
\begin{equation}
g_{\mu\nu}=\begin{pmatrix}
-c^2 & 0 & 0 & 0\\
0 & 1 & 0 & 0\\
0 & 0 & r^2(1+\omega^2) & \omega r\\
0 & 0 & \omega r & 1
\end{pmatrix},
\label{eq:metric-matrix}
\end{equation}
with inverse
\begin{equation}
g^{\mu\nu}=\begin{pmatrix}
-c^{-2} & 0 & 0 & 0\\
0 & 1 & 0 & 0\\
0 & 0 & r^{-2} & -\omega r^{-1}\\
0 & 0 & -\omega r^{-1} & 1+\omega^2
\end{pmatrix}.
\label{eq:inverse-metric}
\end{equation}
The off-diagonal component $g_{\phi z}=\omega r$ encodes the coupling between angular and longitudinal displacements, which is the geometric origin of the helical structure. We note that the metric is stationary, axially symmetric, and translationally invariant along the $z$-direction, but it is not diagonal in the standard cylindrical coordinate basis.

To handle spinor fields in this curved background, we introduce an orthonormal coframe (vierbein) $\{e^a\}$ with $a=0,1,2,3$ labeling the local Lorentz frame. A convenient choice that respects the cylindrical symmetry is
\begin{subequations}
\label{eq:coframe}
\begin{align}
e^0 &= c\,dt, \\
e^1 &= dr, \\
e^2 &= r\,d\phi, \\
e^3 &= dz + \omega r\,d\phi.
\end{align}
\end{subequations}
These one-forms satisfy
\begin{equation}
g_{\mu\nu}=\eta_{ab}\,e^a{}_\mu e^b{}_\nu,
\end{equation}
with $\eta_{ab}=\text{diag}(-1,+1,+1,+1)$ in the mostly-plus signature convention. Equivalently, their inverse satisfies
\begin{equation}
g_{\mu\nu}E_a{}^\mu E_b{}^\nu=\eta_{ab}.
\end{equation}
The components of the coframe and its inverse (the tetrad fields) are
\begin{equation}
e^{a}{}_\mu=\begin{pmatrix}
c & 0 & 0 & 0\\
0 & 1 & 0 & 0\\
0 & 0 & r & 0\\
0 & 0 & \omega r & 1
\end{pmatrix},
\label{eq:coframe-components}
\end{equation}
and
\begin{equation}
E_a{}^{\mu}=\begin{pmatrix}
c^{-1} & 0 & 0 & 0\\
0 & 1 & 0 & 0\\
0 & 0 & r^{-1} & 0\\
0 & 0 & -\omega & 1
\end{pmatrix}.
\label{eq:tetrad-components}
\end{equation}
One can verify that these satisfy $e^a{}_\mu E_b{}^\mu=\delta^a{}_b$ and $E_a{}^\mu e^a{}_\nu=\delta^\mu{}_\nu$, confirming that they form a proper vierbein pair.

The curved-space gamma matrices are constructed from the flat-space Dirac matrices $\gamma^a$ by the standard prescription
\begin{equation}
\gamma^\mu(x)=E_a{}^\mu(x)\gamma^a.
\end{equation}
In the Dirac representation, where
\begin{equation}
\gamma^0=\begin{pmatrix}
\mathbb{I}_2 & 0\\
0 & -\mathbb{I}_2
\end{pmatrix},\quad
\gamma^i=\begin{pmatrix}
0 & \sigma^i\\
-\sigma^i & 0
\end{pmatrix},
\label{eq:flat-gammas}
\end{equation}
with $\sigma^i$ being the Pauli matrices, we obtain the curved gamma matrices explicitly as
\begin{subequations}
\label{eq:curved-gammas}
\begin{align}
\gamma^t &= \frac{1}{c}\,\gamma^0, \\
\gamma^r &= \gamma^1, \\
\gamma^\phi &= \frac{1}{r}\,\gamma^2, \\
\gamma^z &= -\omega\,\gamma^2 + \gamma^3.
\end{align}
\end{subequations}
These satisfy the curved-space anticommutation relations $\{\gamma^\mu,\gamma^\nu\}=-2g^{\mu\nu}\mathbb{I}_4$. The overall sign reflects our conventions: the Dirac-representation matrices adopted here obey the local Clifford algebra $\{\gamma^a,\gamma^b\}=2\,\mathrm{diag}(+1,-1,-1,-1)$, which differs by an overall sign from the mostly-plus frame metric $\eta_{ab}=\mathrm{diag}(-1,+1,+1,+1)$, so that $\gamma^\mu=E_a{}^\mu\gamma^a$ gives $\{\gamma^\mu,\gamma^\nu\}=-2g^{\mu\nu}\mathbb{I}_4$. In particular, the mixed anticommutator involving the angular and longitudinal components yields
\begin{equation}
\{\gamma^\phi,\gamma^z\}=-2g^{\phi z}=+\frac{2\omega}{r}\,\mathbb{I}_4,
\label{eq:key-anticommutator}
\end{equation}
which is nonzero due to the off-diagonal component of the metric. This anticommutator will play a crucial role in generating the geometry-induced Coulomb-like term in the radial equation, as we shall demonstrate explicitly in Section~\ref{sec:decoupling}.

We now turn to the computation of the spin connection. In the torsion-free Levi--Civita formulation, the spin connection one-forms $\omega^a{}_b$ are determined by Cartan's first structure equation
\begin{equation}
de^a + \omega^a{}_b \wedge e^b = 0,
\label{eq:cartan-first}
\end{equation}
where $d$ denotes the exterior derivative and the wedge product is understood. For the coframe~\eqref{eq:coframe}, we compute the exterior derivatives
\begin{subequations}
\begin{align}
de^0 &= 0, \\
de^1 &= 0, \\
de^2 &= dr \wedge d\phi, \\
de^3 &= \omega\,dr \wedge d\phi.
\end{align}
\end{subequations}
Solving Eq.~\eqref{eq:cartan-first} with the antisymmetry condition $\omega_{ab}=-\omega_{ba}$, we find that the nonzero Levi--Civita connection one-forms can be written as
\begin{subequations}
\label{eq:spin-connection-forms}
\begin{align}
\omega^{12} &= -\frac{2+\omega^2}{2}\,d\phi-\frac{\omega}{2r}\,dz, \\
\omega^{13} &= -\frac{\omega}{2}\,d\phi, \\
\omega^{23} &= \frac{\omega}{2r}\,dr.
\end{align}
\end{subequations}
All other components either vanish or are fixed by antisymmetry.

Equivalently, the nonvanishing coordinate components are
\begin{align}
\omega^{2}{}_{3\,r} &= \frac{\omega}{2r}, &
\omega^{1}{}_{2\,\phi} &= -\frac{2+\omega^2}{2}, \nonumber\\
\omega^{1}{}_{3\,\phi} &= -\frac{\omega}{2}, &
\omega^{1}{}_{2\,z} &= -\frac{\omega}{2r}.
\end{align}
together with the corresponding antisymmetric partners. These are the components that enter the spinorial connection.

The spin connection enters the Dirac equation through the spinorial connection matrices $\Omega_\mu^{(\mathrm{LC})}$, defined by
\begin{equation}
\Omega_\mu^{(\mathrm{LC})} = -\frac{1}{8}\,\omega_{\mu ab}\,[\gamma^a,\gamma^b],
\label{eq:spinor-connection-def}
\end{equation}
where the commutator $[\gamma^a,\gamma^b]=\gamma^a\gamma^b-\gamma^b\gamma^a$ generates the Lorentz algebra in the spinor representation. Evaluating this expression using~\eqref{eq:spin-connection-forms} and the flat Dirac matrices, we obtain the $4\times 4$ matrices for each coordinate direction. The time component is identically zero:
\begin{equation}
\Omega_t^{(\mathrm{LC})} = 0.
\label{eq:Omega-t}
\end{equation}
The radial component is
\begin{equation}
\Omega_r^{(\mathrm{LC})} = \frac{i\omega}{4r}
\begin{pmatrix}
0 & 1 & 0 & 0\\
1 & 0 & 0 & 0\\
0 & 0 & 0 & 1\\
0 & 0 & 1 & 0
\end{pmatrix}.
\label{eq:Omega-r}
\end{equation}
The angular component reads
\begin{equation}
\Omega_\phi^{(\mathrm{LC})}=\frac{1}{4}
\begin{pmatrix}
-i a_\omega & \omega & 0 & 0\\
-\omega & i a_\omega & 0 & 0\\
0 & 0 & -i a_\omega & \omega\\
0 & 0 & -\omega & i a_\omega
\end{pmatrix},
\qquad a_\omega\equiv 2+\omega^2 .
\label{eq:Omega-phi}
\end{equation}
and the longitudinal component is
\begin{equation}
\Omega_z^{(\mathrm{LC})} = \frac{i\omega}{4r}
\begin{pmatrix}
-1 & 0 & 0 & 0\\
0 & 1 & 0 & 0\\
0 & 0 & -1 & 0\\
0 & 0 & 0 & 1
\end{pmatrix}.
\label{eq:Omega-z}
\end{equation}

For practical calculations involving two-component spinors, it is convenient to decompose these $4\times 4$ matrices into $2\times 2$ Pauli blocks. In the Dirac representation, where the upper and lower components correspond to particle and antiparticle sectors, respectively, we can write each connection matrix in the block-diagonal form
\begin{equation}
\Omega_\mu^{(\mathrm{LC})} = \begin{pmatrix}
A_\mu & 0\\
0 & B_\mu
\end{pmatrix},
\label{eq:block-structure}
\end{equation}
where $A_\mu$ and $B_\mu$ are $2\times 2$ matrices. For the components computed above, we find
\begin{subequations}
\label{eq:pauli-blocks}
\begin{align}
\Omega_r^{(\mathrm{LC})} &= \frac{i\omega}{4r}\,\mathrm{diag}(\sigma_x,\sigma_x), \\
\Omega_\phi^{(\mathrm{LC})} &= \mathrm{diag}(X_\phi,X_\phi), \\
\Omega_z^{(\mathrm{LC})} &= -\frac{i\omega}{4r}\,\mathrm{diag}(\sigma_z,\sigma_z),
\end{align}
\end{subequations}
where we have introduced the compact notation
\begin{equation}
X_\phi = -\frac{i}{4}(2+\omega^2)\,\sigma_z + \frac{i\omega}{4}\,\sigma_y.
\label{eq:X-phi-def}
\end{equation}
The block-diagonal structure reflects the fact that, in the absence of external fields coupling the particle and antiparticle sectors, the spin connection acts separately on each sector.

The explicit form of these connection matrices will be essential for our subsequent analysis. In particular, when we construct the first-order Dirac system in Section~\ref{sec:dirac-oscillator}, these components combine with the curved gamma matrices to produce the effective spinorial shift operator $\hat{\mathcal{S}}_{\mathrm{LC}}(r)$ that appears in the off-diagonal blocks of the Hamiltonian. The combination $i\hbar c\,\gamma^\mu\Omega_\mu^{(\mathrm{LC})}$ is evaluated explicitly in Sec.~\ref{sec:dirac-oscillator}, where it supplies both the standard cylindrical $1/(2r)$ contribution and the helical scalar shift needed for the correct radial operator.

\section{Axial torsion and contortion}
\label{sec:torsion}

The Levi--Civita connection computed in the previous section is entirely determined by the metric and corresponds to a torsion-free geometry. However, in more general formulations of gravitational theories, particularly in the Einstein--Cartan framework~\cite{AP.1961.12.2.200,ARXIV.2006.gr-qc.0606062}, spacetime can possess intrinsic torsion that couples to the spin density of matter fields. Torsion manifests itself through the contortion tensor $K^\lambda{}_{\mu\nu}$, which is related to the torsion tensor $T^\lambda{}_{\mu\nu}$ by a specific algebraic relation. The full spin connection then decomposes as $\Omega_\mu = \Omega_\mu^{(\mathrm{LC})} + \Omega_\mu^{(\mathrm{cont})}$, where $\Omega_\mu^{(\mathrm{cont})}$ encodes the contribution from torsion.

From a physical perspective, torsion in Einstein--Cartan theory arises as a response to fermionic spin currents, analogous to how curvature responds to energy-momentum. In condensed matter systems, effective torsion can emerge in materials with topological defects~\cite{AP.1992.216.1.1,IJTP.1987.26.8.715} or in models of quantum field theory on structured backgrounds~\cite{WS.1989.0356}. The mathematical structure of torsion admits a decomposition into irreducible components under the Lorentz group: vector (or trace), axial-vector (or pseudotrace), and purely tensorial parts. Each component couples differently to fermion fields and has distinct physical interpretations. For our purposes, we adopt a phenomenological approach by introducing a purely axial-torsion configuration, which is among the simplest and most physically motivated scenarios. Axial torsion couples exclusively to the axial current $\bar{\psi}\gamma^\mu\gamma^5\psi$ of fermions and has been studied in various contexts, including chiral anomalies~\cite{PRD.2018.98.025016}, neutrino physics, and effective field theories on topological materials.

A purely axial contortion is characterized by a pseudovector field $S^\mu(x)$ that defines the torsion structure. At the level of the Dirac equation, this axial coupling enters through an additional term in the covariant derivative acting on the spinor field. We parameterize this contribution through
\begin{equation}
i\hbar c\,\gamma^\mu\Omega_\mu^{(\mathrm{cont})} 
= \lambda_{\mathrm{ax}}\,\hbar c\,S^0\,\gamma^0\gamma^5,
\label{eq:axial-prescription}
\end{equation}
where $\gamma^5=i\gamma^0\gamma^1\gamma^2\gamma^3$ is the chirality matrix and $\lambda_{\mathrm{ax}}$ is a dimensionless coupling constant. In the standard Einstein--Cartan formulation with minimal coupling to fermions, the coefficient takes the value $\lambda_{\mathrm{ax}}=3/4$, although we shall keep it arbitrary to allow for generalizations to effective theories or modified gravity scenarios.

We choose the pseudovector to be aligned with the timelike direction and constant throughout spacetime:
\begin{equation}
S^\mu = (S^0, 0, 0, 0),
\label{eq:S-choice}
\end{equation}
where $S^0$ is a constant with dimensions of inverse length. This choice of a purely axial torsion configuration is motivated by several physical and mathematical considerations. First, it preserves all the symmetries of the helically twisted metric~\eqref{eq:metric}: stationarity under time translations, axial symmetry under rotations, and translational invariance along the longitudinal direction. Any torsion configuration with spatial components would necessarily break one or more of these symmetries, complicating the analysis and obscuring the interplay between geometric and torsional effects. Second, in the Einstein--Cartan framework, torsion is sourced by the spin density of matter through the relation $T^\lambda{}_{\mu\nu} \propto s^\lambda{}_{\mu\nu}$, where $s^\lambda{}_{\mu\nu}$ is the spin density tensor. The configuration~\eqref{eq:S-choice} corresponds to a uniform, time-independent distribution of spin density aligned along the temporal direction, analogous to a background of polarized fermions with their spins oriented in the rest frame. This represents a physically realizable scenario in systems with macroscopic spin polarization or in effective field theories where torsion parameterizes collective spin effects. Third, the axial nature of this coupling means it distinguishes left-handed from right-handed fermions through the chirality matrix $\gamma^5$. Under parity transformations, $\gamma^5$ changes sign while $S^0$ remains invariant as a pseudoscalar, making this configuration relevant for studying chiral phenomena, parity-violating processes, and anomalies in curved spacetime. Finally, this choice represents the simplest nontrivial extension beyond the purely geometric torsion-free case while still allowing for analytical treatment and clear physical interpretation of the results.

In the Dirac basis, the chirality matrix has the explicit block form
\begin{equation}
\gamma^5 = \begin{pmatrix}
0 & \mathbb{I}_2\\
\mathbb{I}_2 & 0
\end{pmatrix},
\label{eq:gamma5}
\end{equation}
from which we immediately obtain
\begin{equation}
\gamma^0\gamma^5 = \begin{pmatrix}
0 & \mathbb{I}_2\\
-\mathbb{I}_2 & 0
\end{pmatrix}.
\label{eq:gamma0-gamma5}
\end{equation}
Substituting Eqs.~\eqref{eq:S-choice} and~\eqref{eq:gamma0-gamma5} into the prescription~\eqref{eq:axial-prescription}, we find
\begin{equation}
i\hbar c\,\gamma^\mu\Omega_\mu^{(\mathrm{cont})} 
= \lambda_{\mathrm{ax}}\,\hbar c\,S^0
\begin{pmatrix}
0 & \mathbb{I}_2\\
-\mathbb{I}_2 & 0
\end{pmatrix}.
\label{eq:axial-block-full}
\end{equation}

This block structure reveals an important feature of axial torsion: it enters the Dirac equation only in the off-diagonal blocks that couple the upper and lower two-component spinors. In contrast to the Levi--Civita connection, which contributes nontrivial $2\times 2$ matrix structures involving Pauli matrices, the axial contortion contributes as a simple scalar multiple of the identity matrix in each off-diagonal block. When we separate the four-component Dirac equation into a coupled system for the upper and lower two-component spinors $\psi_A$ and $\psi_B$, this axial contribution manifests as an additive constant shift in the effective operators coupling these components.

Specifically, in the first-order Dirac system that we shall derive in the next section, the off-diagonal operator acting between $\psi_A$ and $\psi_B$ will contain both kinetic terms, geometric contributions from the Levi--Civita connection, and the torsional shift. We define the axial spinorial shift operator as
\begin{equation}
\hat{\mathcal{S}}_{\mathrm{ax}} 
= \lambda_{\mathrm{ax}}\,\hbar c\,S^0\,\mathbb{I}_2,
\label{eq:S-ax-def}
\end{equation}
which is one of the two fundamental shift operators, along with $\hat{\mathcal{S}}_{\mathrm{LC}}(r)$ from the Levi--Civita sector, that will appear explicitly in our formulation. The operator $\hat{\mathcal{S}}_{\mathrm{ax}}$ is constant in both space and spinor indices, reflecting the simplicity of the uniform axial torsion configuration we have chosen.

Several physical aspects of this construction deserve emphasis. The sign difference between the upper-right and lower-left blocks in Eq.~\eqref{eq:axial-block-full} reflects the pseudoscalar nature of the axial coupling: under parity transformations, $\gamma^5$ changes sign while $\gamma^0$ does not, leading to opposite contributions for particle and antiparticle sectors. The constant $S^0$ can be interpreted physically in multiple ways depending on the context. In Einstein--Cartan theory, it represents the magnitude of the background torsion density sourced by spin polarization. In effective field theories of condensed matter systems, it may parameterize the strength of spin-orbit coupling induced by lattice defects or strain. In phenomenological models of modified gravity, it could quantify deviations from general relativity due to torsional degrees of freedom. In all cases, $S^0$ would ideally be determined dynamically by solving the field equations with appropriate boundary conditions and matter sources, but in our phenomenological approach, we treat it as an external parameter whose value characterizes the strength of torsion. When $S^0=0$, the formalism reduces exactly to the purely geometric case with only the Levi--Civita connection, allowing for a clean separation between metric and torsion effects and facilitating the identification of purely torsional contributions to observable quantities.

The decomposition of the full spin connection into $\Omega_\mu = \Omega_\mu^{(\mathrm{LC})} + \Omega_\mu^{(\mathrm{cont})}$ ensures that all geometric and torsional contributions are systematically accounted for in the Dirac equation. In the next section, we shall incorporate both $\hat{\mathcal{S}}_{\mathrm{LC}}(r)$ and $\hat{\mathcal{S}}_{\mathrm{ax}}$ into the first-order Dirac oscillator system, demonstrating how these shift operators combine with the kinetic terms and the Moshinsky coupling to produce the complete effective Hamiltonian.

\section{Dirac oscillator: first-order system}
\label{sec:dirac-oscillator}

We now formulate the Dirac oscillator in the helically twisted spacetime, keeping both the Levi--Civita and axial torsion contributions. The starting point is the curved-space Dirac equation
\begin{equation}
\left[i\hbar c\,\gamma^\mu(\partial_\mu+\Omega_\mu)-Mc^2\right]\Psi=0,
\label{eq:dirac-curved-corrected}
\end{equation}
where
\begin{equation}
\Omega_\mu=\Omega_\mu^{(\mathrm{LC})}+\Omega_\mu^{(\mathrm{cont})}.
\end{equation}
The Dirac oscillator is introduced through the standard Moshinsky prescription
\begin{equation}
\boldsymbol{p}\longrightarrow \boldsymbol{p}-iM\omega_D\beta\,\boldsymbol{r},
\qquad
\beta=\gamma^0.
\label{eq:moshinsky-prescription-corrected}
\end{equation}
Since the oscillator considered here is radial, the coupling is implemented along the local radial direction of the orthonormal frame. Thus, the Dirac equation becomes
\begin{equation}
\left[i\hbar c\,\gamma^\mu(\partial_\mu+\Omega_\mu)-Mc^2
+iM\omega_Dc\,\gamma^1\beta r\right]\Psi=0.
\label{eq:dirac-oscillator-beta}
\end{equation}
The presence of $\beta$ is, of course, part of the standard Moshinsky prescription. What is essential here is that it changes the sign of the oscillator term in the two Pauli blocks; consequently, after decoupling, the second-order equation is not obtained by squaring a single operator.

The spacetime is stationary, axially symmetric, and invariant under translations along $z$. We therefore use the separated form
\begin{equation}
\Psi(t,r,\phi,z)=e^{-iEt/\hbar}e^{im\phi}e^{ikz}
\begin{pmatrix}
\psi_A(r)\\[2pt]
\psi_B(r)
\end{pmatrix},
\label{eq:separation-ansatz-corrected}
\end{equation}
where $E$ is the energy eigenvalue, $k$ is the longitudinal wave number, and $m$ denotes the eigenvalue associated with the angular dependence in this representation. When one compares the flat cylindrical limit with the usual planar notation, it is convenient to write $m=l+1/2$, with $l\in\mathbb{Z}$.

Using the curved gamma matrices
\begin{equation}
\gamma^r=\gamma^1,
\qquad
\gamma^\phi=\frac{1}{r}\gamma^2,
\qquad
\gamma^z=-\omega\gamma^2+\gamma^3,
\end{equation}
the angular and longitudinal derivatives give
\begin{equation}
\partial_\phi\Psi=im\Psi,
\qquad
\partial_z\Psi=ik\Psi.
\end{equation}
The corresponding momentum-like quantities in the two-component system are
\begin{equation}
P_\phi=\hbar\left(\frac{m}{r}-\omega k\right),
\qquad
P_z=\hbar k.
\label{eq:Pphi-Pz-corrected}
\end{equation}
The combination $m/r-\omega k$ is the first explicit manifestation of the helical geometry. It shows that angular and longitudinal motion are not dynamically independent in this background.

The contribution of the Levi--Civita spin connection to the off-diagonal blocks is
\begin{equation}
i\hbar c\,\gamma^\mu\Omega^{(\mathrm{LC})}_\mu
=
\begin{pmatrix}
0&B_{\mathrm{LC}}\\
-B_{\mathrm{LC}}&0
\end{pmatrix},
\end{equation}
with
\begin{equation}
B_{\mathrm{LC}}=
\hbar c\left(\frac{i}{2r}\sigma_x-\frac{\omega}{4r}\mathbb{I}_2\right).
\label{eq:BLC-corrected}
\end{equation}
Similarly, the axial torsion contribution is
\begin{equation}
i\hbar c\,\gamma^\mu\Omega^{(\mathrm{cont})}_\mu
=
\lambda_{\mathrm{ax}}\hbar cS^0
\begin{pmatrix}
0&\mathbb{I}_2\\
-\mathbb{I}_2&0
\end{pmatrix}.
\label{eq:Bax-corrected}
\end{equation}
When the off-diagonal terms are moved to the right-hand side of the two coupled equations, the scalar part appears with the sign
\begin{equation}
q(r)=\frac{\hbar c\omega}{4r}-\lambda_{\mathrm{ax}}\hbar cS^0.
\label{eq:q-corrected}
\end{equation}
This convention will be used throughout the decoupling calculation below.

It remains to organize the radial operator. In cylindrical coordinates, the spin connection supplies the standard $1/(2r)$ contribution. We define
\begin{equation}
D_r\equiv \frac{d}{dr}+\frac{1}{2r},
\qquad
p_D\equiv -i\hbar D_r.
\label{eq:pD-corrected}
\end{equation}
The radial Moshinsky term has opposite signs in the two Pauli blocks. Hence we introduce
\begin{equation}
A_+\equiv p_D+iM\omega_D r,
\qquad
A_-\equiv p_D-iM\omega_D r.
\label{eq:Apm-corrected}
\end{equation}
With these definitions, the Dirac equation reduces to the coupled first-order system
\begin{subequations}
\label{eq:first-order-corrected}
\begin{align}
(E-Mc^2)\psi_A&=\hat\Pi_+\psi_B,\\
(E+Mc^2)\psi_B&=\hat\Pi_-\psi_A,
\end{align}
\end{subequations}
where
\begin{equation}
\hat\Pi_+=c\left(\sigma_xA_+ +\sigma_yP_\phi+\sigma_zP_z\right)+q(r)\mathbb{I}_2,
\label{eq:Pi-plus-corrected}
\end{equation}
\begin{equation}
\hat\Pi_-=c\left(\sigma_xA_- +\sigma_yP_\phi+\sigma_zP_z\right)+q(r)\mathbb{I}_2.
\label{eq:Pi-minus-corrected}
\end{equation}
Equations~\eqref{eq:first-order-corrected}--\eqref{eq:Pi-minus-corrected} are the correct first-order formulation of the standard Dirac oscillator in the helically twisted spacetime. The operators $\hat\Pi_+$ and $\hat\Pi_-$ are not equal because the oscillator coupling contains $\beta$.

\section{Decoupling and radial matrix equation}
\label{sec:decoupling}

The upper component is obtained by eliminating $\psi_B$ from Eq.~\eqref{eq:first-order-corrected}. Applying $\hat\Pi_+$ to the second equation or, equivalently, substituting
\begin{equation}
\psi_B=\frac{1}{E+Mc^2}\hat\Pi_-\psi_A
\end{equation}
into the first equation, we find
\begin{equation}
(E^2-M^2c^4)\psi_A=\hat\Pi_+\hat\Pi_-\psi_A.
\label{eq:second-order-corrected-master}
\end{equation}
For the lower component, one analogously obtains
\begin{equation}
(E^2-M^2c^4)\psi_B=\hat\Pi_-\hat\Pi_+\psi_B.
\end{equation}
Therefore, the second-order operator is an ordered product, not the square of a single operator.

Let
\begin{equation}
\begin{aligned}
K_+&\equiv \sigma_xA_+ +\sigma_yP_\phi+\sigma_zP_z,\\
K_-&\equiv \sigma_xA_- +\sigma_yP_\phi+\sigma_zP_z.
\end{aligned}
\end{equation}
Then
\begin{equation}
\hat\Pi_+\hat\Pi_-
=c^2K_+K_-+c(K_+q+qK_-)+q^2\mathbb{I}_2.
\label{eq:PiPi-corrected}
\end{equation}
The mixed term is not the anticommutator $c\{K,q\}$, because $K_+\neq K_-$.

\subsection{Evaluation of $K_+K_-$}

Using
\begin{equation}
\sigma_i\sigma_j=\delta_{ij}\mathbb{I}_2+i\epsilon_{ijk}\sigma_k,
\end{equation}
one obtains
\begin{align}
K_+K_-={}&
-\hbar^2\frac{d^2}{dr^2}
-\frac{\hbar^2}{r}\frac{d}{dr}
+\frac{\hbar^2}{4r^2}
\nonumber\\
&+M^2\omega_D^2r^2-M\hbar\omega_D
+P_\phi^2+P_z^2
\nonumber\\
&-\left(\frac{\hbar^2m}{r^2}+2M\omega_DrP_\phi\right)\sigma_z
+2M\omega_DrP_z\sigma_y.
\label{eq:KpKm-final-corrected}
\end{align}

\subsection{Terms involving the scalar shift $q(r)$}

We now compute the terms containing $q(r)$. Since $q$ is a scalar function of $r$,
\begin{equation}
K_+q+qK_-=
\sigma_x(A_+q+qA_-)+2qP_\phi\sigma_y+2qP_z\sigma_z.
\label{eq:Kq-corrected-start}
\end{equation}
The oscillator terms in $A_+q+qA_-$ cancel because $q(r)$ commutes with $r$. Hence
\begin{equation}
A_+q+qA_-=p_Dq+qp_D=\{p_D,q\}.
\end{equation}
Acting on a radial function $f(r)$,
\begin{align}
\{p_D,q\}f
&=-i\hbar D_r(qf)-i\hbar qD_rf
\nonumber\\
&=-i\hbar\left(2q\frac{df}{dr}+q'f+\frac{q}{r}f\right).
\end{align}
Therefore
\begin{equation}
\{p_D,q\}=-i\hbar\left(2q\frac{d}{dr}+q'+\frac{q}{r}\right).
\label{eq:pdq-corrected}
\end{equation}
For
\begin{equation}
q(r)=\frac{\hbar c\omega}{4r}-\lambda_{\mathrm{ax}}\hbar cS^0,
\end{equation}
we have
\begin{equation}
q'(r)=-\frac{\hbar c\omega}{4r^2}.
\end{equation}
Thus
\begin{equation}
K_+q+qK_-
=\sigma_x\{p_D,q\}+2qP_\phi\sigma_y+2qP_z\sigma_z.
\label{eq:Kq-final-corrected}
\end{equation}

\subsection{Exact radial matrix equation}

Substituting Eqs.~\eqref{eq:KpKm-final-corrected} and~\eqref{eq:Kq-final-corrected} into Eq.~\eqref{eq:PiPi-corrected}, the equation for the upper two-component spinor becomes
%\begin{widetext}
\begin{align}
&\Bigg\{
 c^2\Big[
-\hbar^2\left(\frac{d^2}{dr^2}+\frac{1}{r}\frac{d}{dr}\right)
+\frac{\hbar^2}{4r^2}
+M^2\omega_D^2r^2
\nonumber\\
&-M\hbar\omega_D
+P_\phi^2+P_z^2
\Big]
+q^2-(E^2-M^2c^4)
\Bigg\}\psi_A
\nonumber\\
&+c^2\left[
-\left(\frac{\hbar^2m}{r^2}+2M\omega_DrP_\phi\right)\sigma_z
+2M\omega_DrP_z\sigma_y
\right]\psi_A
\nonumber\\
&+c\bigg[
-i\hbar\left(2q\frac{d}{dr}+q'+\frac{q}{r}\right)\sigma_x
\nonumber\\
&\hspace{1.6cm}
+2qP_\phi\sigma_y
+2qP_z\sigma_z
\bigg]\psi_A=0.
\label{eq:radial-matrix-final-corrected}
\end{align}
%\end{widetext}
Here
\begin{equation}
\begin{aligned}
&P_\phi=\hbar\left(\frac{m}{r}-\omega k\right),
\qquad
P_z=\hbar k,\\
&q(r)=\frac{\hbar c\omega}{4r}-\lambda_{\mathrm{ax}}\hbar cS^0.
\end{aligned}
\end{equation}
Equation~\eqref{eq:radial-matrix-final-corrected} is the exact radial equation, up to this point, for the standard radial Dirac oscillator in the helically twisted background with axial torsion. It is a matrix equation in the two-component spin space. In particular, when $k\neq0$, the term proportional to $\sigma_y$ remains present. This is not an inconsistency; rather, it reflects the fact that the oscillator is radial while the helical defect couples angular and longitudinal motion. Therefore, the problem cannot, in general, be reduced to a scalar radial equation by a direct projection onto eigenstates of $\sigma_z$.

\subsection{Explicit coupled radial system}
\label{sec:coupled-system}

To exhibit Eq.~\eqref{eq:radial-matrix-final-corrected} in its final form, we
insert the separated solution~\eqref{eq:separation-ansatz-corrected}, for which
the upper spinor is $\psi_A=(\psi_{A,1},\psi_{A,2})^{\mathsf T}$, together with the
explicit momentum-like quantities~\eqref{eq:Pphi-Pz-corrected} and the scalar
shift~\eqref{eq:q-corrected}, $P_\phi=\hbar(m/r-\omega k)$, $P_z=\hbar k$, and
$q(r)=\hbar c\omega/4r-\lambda_{\mathrm{ax}}\hbar cS^0$. Carrying out the
substitution and writing the two spinor components, the matrix
equation~\eqref{eq:radial-matrix-final-corrected} becomes the coupled system of
two second-order radial equations
\begin{subequations}
\label{eq:coupled-system}
\begin{align}
\big[\hat{\mathcal H}_0+V(r)+Z(r)\big]\psi_{A,1}
+\hat{\mathcal C}(r)\,\psi_{A,2}&=\Lambda\,\psi_{A,1},\\
\hat{\mathcal C}^{\dagger}(r)\,\psi_{A,1}
+\big[\hat{\mathcal H}_0+V(r)-Z(r)\big]\psi_{A,2}&=\Lambda\,\psi_{A,2},
\end{align}
\end{subequations}
where $\Lambda\equiv E^2-M^2c^4$ is the spectral parameter and
\begin{equation}
\hat{\mathcal H}_0=-\hbar^2c^2\left(\frac{d^2}{dr^2}+\frac{1}{r}\frac{d}{dr}\right)
\end{equation}
is the common radial operator. The diagonal scalar potential and the $\sigma_z$
term are
\begin{align}
V(r)={}&c^2\bigg[
\frac{\hbar^2\!\left(m^2+\tfrac14+\tfrac{\omega^2}{16}\right)}{r^2}
-\frac{\hbar^2\omega\,(4km+\lambda_{\mathrm{ax}}S^0)}{2r}
\nonumber\\
&+M^2\omega_D^2r^2-M\hbar\omega_D
+\hbar^2(1+\omega^2)k^2
\nonumber\\
&+\hbar^2\lambda_{\mathrm{ax}}^2(S^0)^2\bigg],
\label{eq:Veff}
\\
Z(r)={}&c^2\bigg[
-\frac{\hbar^2m}{r^2}
+\frac{\hbar^2k\omega}{2r}
+2M\hbar\omega_D\,\omega k\,r
\nonumber\\
&-2M\hbar\omega_D m
-2\lambda_{\mathrm{ax}}\hbar^2kS^0\bigg].
\label{eq:Zeff}
\end{align}
The off-diagonal coupling $\hat{\mathcal C}(r)=\hat X(r)-i\,Y(r)$ combines a
first-order (derivative) operator $\hat X$, coming from the $\sigma_x$ scalar
shift, and the multiplicative $\sigma_y$ term $Y$,
\begin{align}
\hat X(r)={}&-i\hbar^2c^2\left[\left(\frac{\omega}{2r}-2\lambda_{\mathrm{ax}}S^0\right)
\frac{d}{dr}-\frac{\lambda_{\mathrm{ax}}S^0}{r}\right],
\label{eq:Xcoupl}
\\
Y(r)={}&c^2\bigg[\frac{\hbar^2m\omega}{2r^2}-\frac{\hbar^2k\omega^2}{2r}
-\frac{2\lambda_{\mathrm{ax}}\hbar^2mS^0}{r}
\nonumber\\
&+2M\hbar\omega_Dk\,r+2\lambda_{\mathrm{ax}}\hbar^2k\omega S^0\bigg].
\label{eq:Ycoupl}
\end{align}
Since $\hat X$ is Hermitian under the radial measure $r\,dr$ and $Y$ is real, the two off-diagonal entries are Hermitian conjugates, $\hat{\mathcal C}^{\dagger}=\hat X+i\,Y$, so that the system is self-adjoint and the eigenvalues $\Lambda$ are real.

System~\eqref{eq:coupled-system} is manifestly non-diagonal: the radial functions
$\psi_{A,1}$ and $\psi_{A,2}$ are coupled by $\hat{\mathcal C}(r)$. The derivative
part $\hat X$ is controlled by the helical twist $\omega$ and the axial torsion
$\lambda_{\mathrm{ax}}S^0$, while the multiplicative part $Y$ is controlled by the
longitudinal momentum $k$. The coupling vanishes only in the limit $\omega\to0$,
$S^0\to0$, $k=0$, where the two equations decouple and reduce to the planar Dirac
oscillator. For generic helical parameters, the two components are intertwined, and
the problem must be solved as the coupled system~\eqref{eq:coupled-system}, as
described in the next section.

\section{Numerical solution of the radial equation}
\label{sec:numerical}

Equation~\eqref{eq:radial-matrix-final-corrected} can be cast as a linear
eigenvalue problem. Since the energy enters only through $E^2$, we define
$\Lambda\equiv E^2-M^2c^4$ and write
\begin{equation}
\hat{H}\,\psi_A=\Lambda\,\psi_A,
\label{eq:eigenproblem}
\end{equation}
where $\hat{H}$ collects all the $r$-dependent operators on the left-hand side
of Eq.~\eqref{eq:radial-matrix-final-corrected}; explicitly,
$\hat H\psi_A=\Lambda\psi_A$ is the coupled two-component
system~\eqref{eq:coupled-system}. The physical energies follow from
$E=\sqrt{M^2c^4+\Lambda}$ on the positive-energy branch. Because the system is
non-diagonal, the two radial components $\psi_{A,1}$ and $\psi_{A,2}$ are solved
simultaneously.

We solve Eq.~\eqref{eq:eigenproblem} with a linear finite-element discretization that respects the radial measure $r\,dr$: the kinetic operator is represented by the weak form $\hbar^2c^2\!\int\psi_A'^{\dagger}\psi_A'\,r\,dr$, while the multiplicative and Pauli terms are integrated by Gauss quadrature. The derivative coupling $\mathcal{L}_x=c\{p_D,q\}$ is assembled in its symmetric form, which guarantees that the discrete $\hat{H}$ is Hermitian and the eigenvalues $\Lambda$ are real. The node at $r=0$ is left free (natural Neumann condition), which is essential to reproduce the $\nu=0$ sector in which $\psi_A$ tends to a
constant at the origin, whereas the centrifugal terms $\propto 1/r^2$ enforce
$\psi_A(0)=0$ automatically whenever the effective angular barrier is nonzero; at
the outer edge we impose $\psi_A(r_{\max})=0$. All results below use natural
units $\hbar=c=M=1$, so that lengths are measured in Compton units $\hbar/Mc$ and
energies in units of $Mc^2$; in this convention $k$ and $\lambda_{\mathrm{ax}}S^0$
are measured in inverse Compton length, whereas the helical parameter $\omega$
remains dimensionless.

As a stringent check, in the flat limit ($\omega\to0$, $S^0\to0$, $k=0$) the
matrix equation decouples into eigenstates of $\sigma_z$ and reproduces the
planar Dirac oscillator, whose spectrum is known in closed form,
\begin{align}
E^2={}&M^2c^4
+2M\hbar c^2\omega_D\bigl(2n+|\nu_s|-s\,\nu_s\bigr),
\nonumber\\
\nu_s={}&l+\frac12(1-s).
\label{eq:analytic-flat}
\end{align}
with $n=0,1,2,\dots$ and $s=\pm1$. The numerical eigenvalues reproduce
Eq.~\eqref{eq:analytic-flat} to a maximum relative error of $\sim5\times10^{-5}$
for several values of $\omega_D$ and $m$, and the five lowest levels are
converged to $\sim10^{-5}$ under mesh refinement. Unless stated otherwise, the
figures below use the reference set $\omega_D=1$, $\omega=0.3$, $k=0.5$,
$m=1/2$, and $\lambda_{\mathrm{ax}}S^0=0$.

\begin{figure}[t]
\centering
\includegraphics[width=\columnwidth]{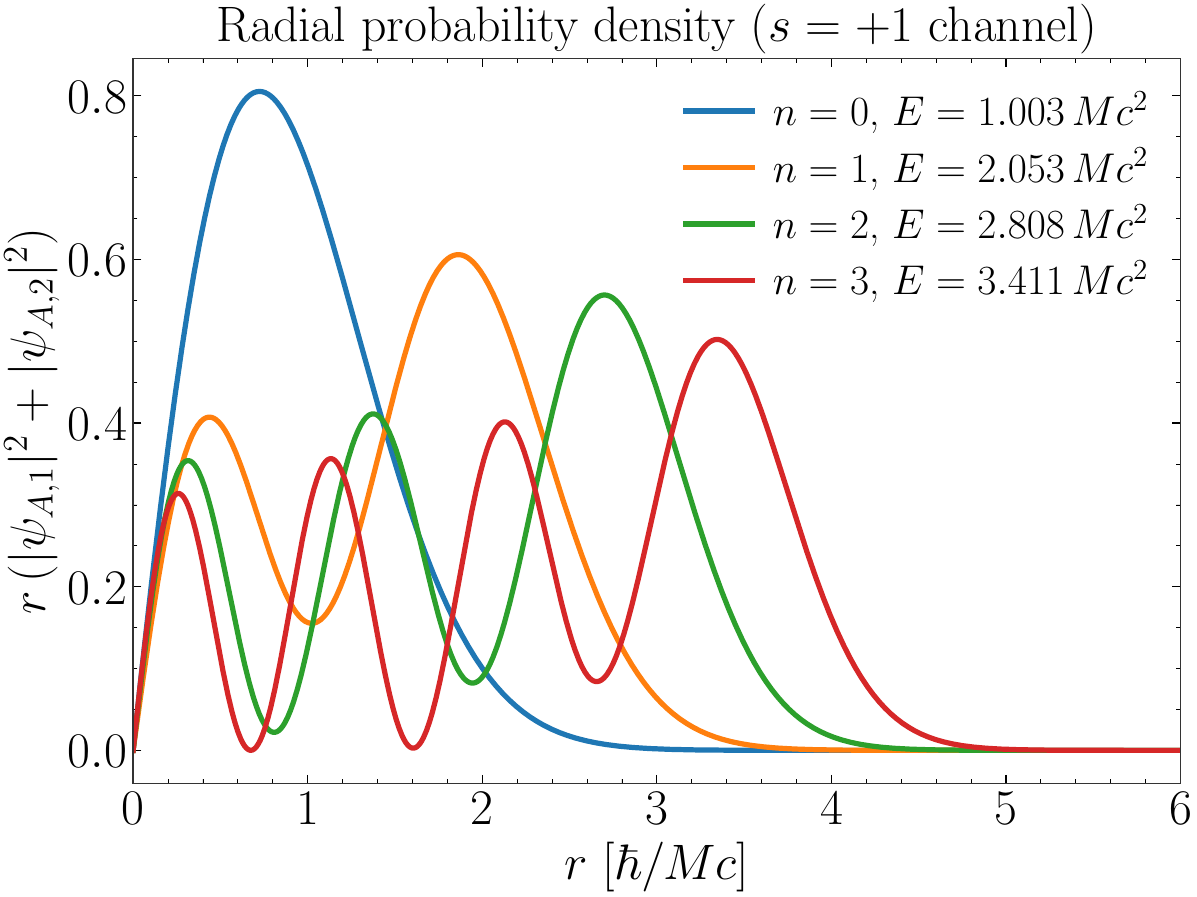}
\caption{Radial probability density
$r\,\bigl(|\psi_{A,1}|^2+|\psi_{A,2}|^2\bigr)$ for the four lowest states of the
$s=+1$ spin sector ($\psi_{A,1}$ dominant), labeled by the radial quantum number
$n$ and computed for the reference parameters. The weight $r$ corresponds to the
two-dimensional radial measure $r\,dr$, so that the area under each curve is
normalized to unity. The $n$-th state exhibits $n+1$ maxima, as expected for
harmonic radial confinement. The full low-energy spectrum interleaves this ladder
with that of the $s=-1$ sector; see Table~\ref{tab:states}.}
\label{fig:density}
\end{figure}

The radial probability densities are shown in Fig.~\ref{fig:density}. The density of the ground state is concentrated within a few Compton lengths of the axis, and the excited states spread outward while developing an increasing number of radial nodes, consistent with the harmonic confinement; the $n$-th state has $n+1$ maxima. The smooth behavior at the origin confirms that the free-node treatment of $r=0$ captures the correct boundary behavior of the $\nu=0$ sector without introducing spurious artifacts. Although only the total density
$|\psi_{A,1}|^2+|\psi_{A,2}|^2$ is displayed, both spin components contribute. For the reference parameters, the first component dominates the $s=+1$ sector, while the second component is generated by the off-diagonal couplings in Eq.~\eqref{eq:coupled-system}; its relative weight increases as the helical twist $\omega$ and the longitudinal momentum $k$ strengthen the angular--longitudinal mixing.
\begin{figure*}[tbhp]
\centering
\includegraphics[width=0.48\textwidth]{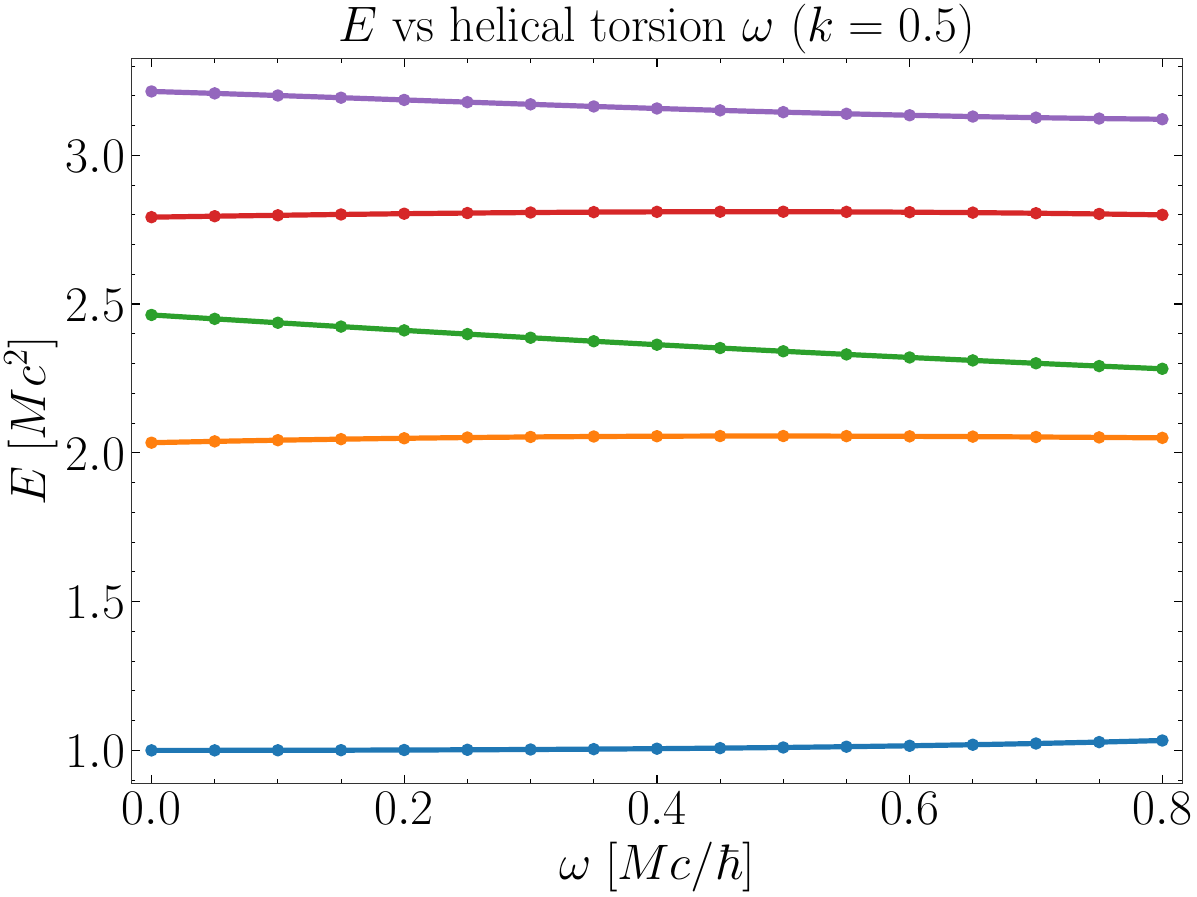}\hfill
\includegraphics[width=0.48\textwidth]{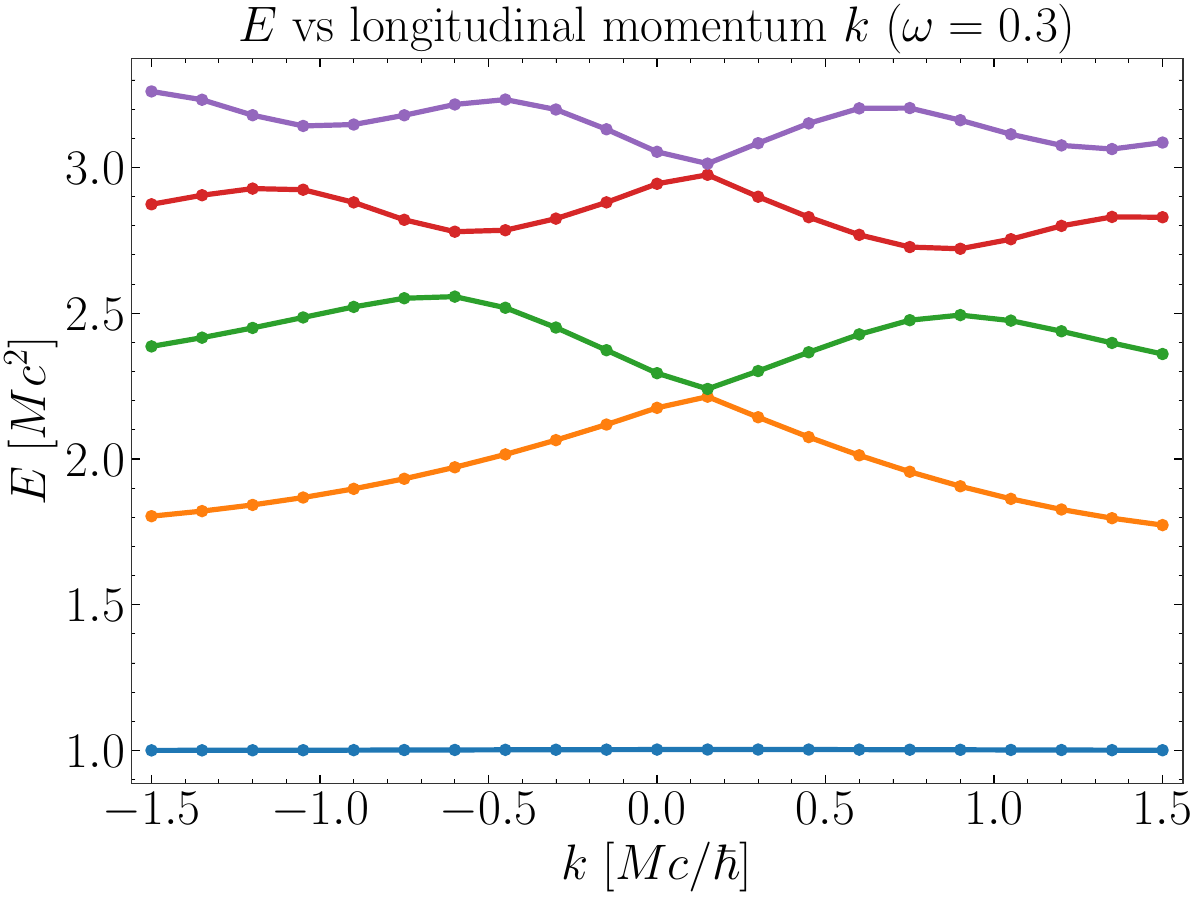}\\[6pt]
\includegraphics[width=0.48\textwidth]{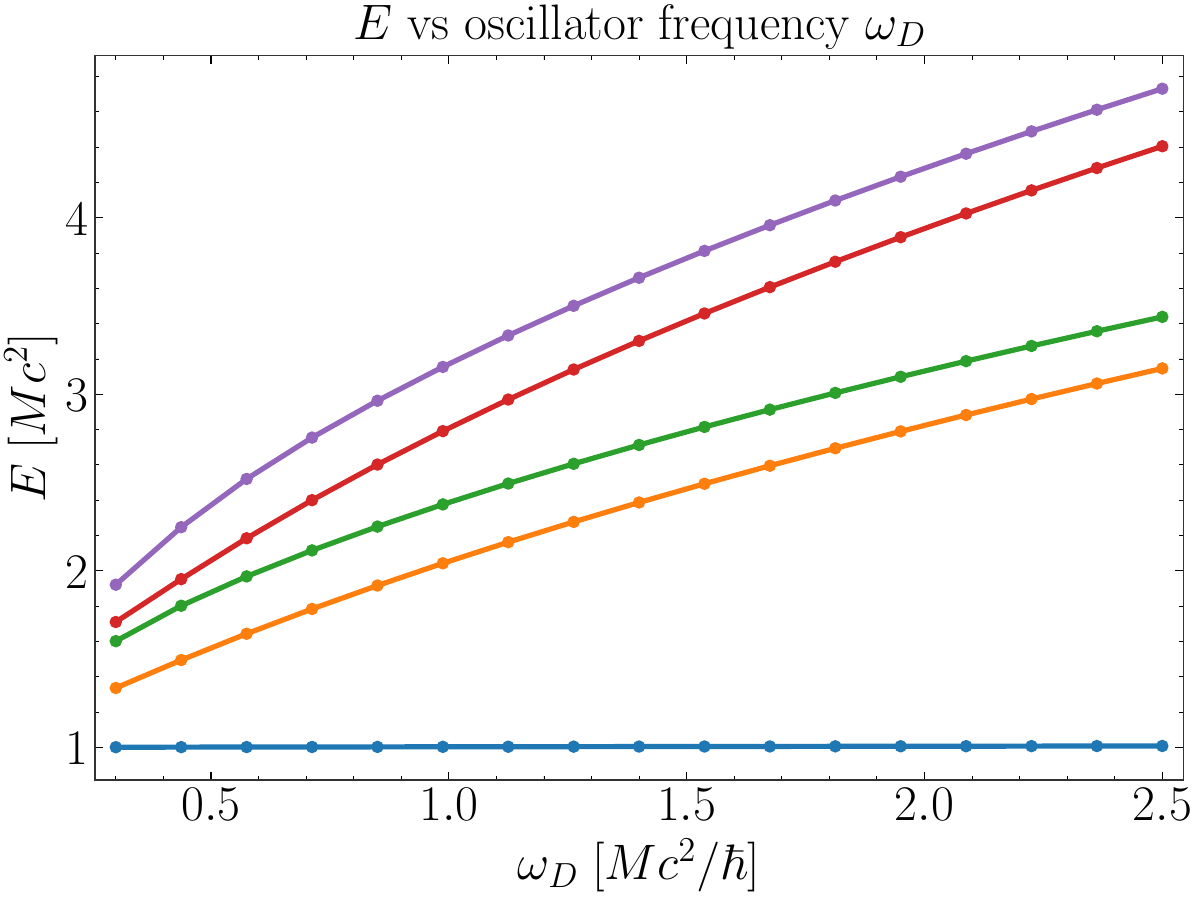}\hfill
\includegraphics[width=0.48\textwidth]{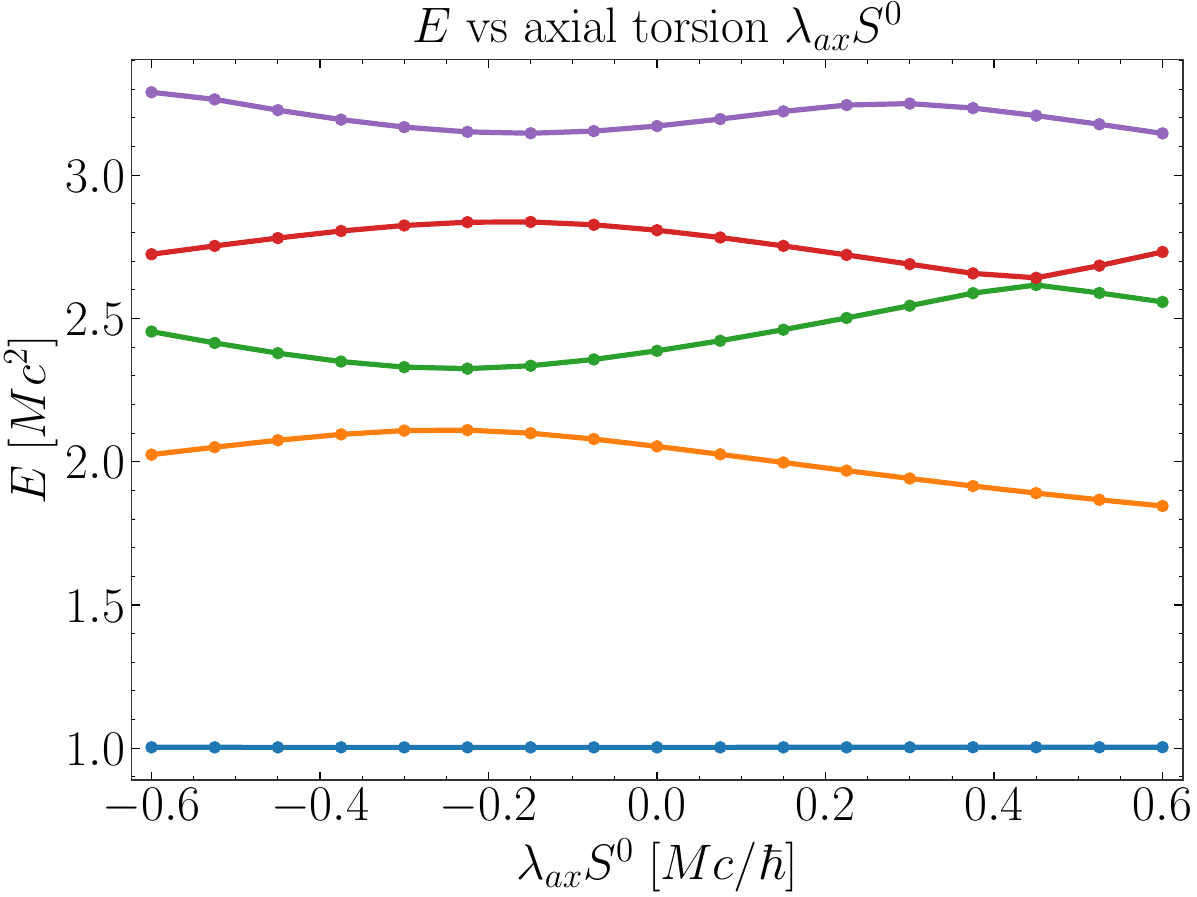}
\caption{Energy spectrum of the five lowest bound states as a function of the
relevant model parameters, with $m=1/2$ and the reference values
$\omega_D=1$, $\omega=0.3$, $k=0.5$, $\lambda_{\mathrm{ax}}S^0=0$ held fixed
except for the parameter being scanned. (a) $E$ versus the helical twist
$\omega$. (b) $E$ versus the longitudinal momentum $k$; the marked asymmetry in
$k$ and the avoided crossings are the spectral fingerprint of the
angular--longitudinal mixing $P_\phi=\hbar(m/r-\omega k)$ enforced by the
off-diagonal metric component. (c) $E$ versus the oscillator frequency
$\omega_D$, showing the monotonic increase of the levels with the strength of the
harmonic confinement. (d) $E$ versus the axial torsion $\lambda_{\mathrm{ax}}S^0$.
The lowest level stays essentially at the rest energy $E\simeq Mc^2$ throughout;
the small offset and its weak rise along the $\omega$ scan in panel~(a) are the
signature of the soft supersymmetry breaking by the helical twist discussed in
Sec.~\ref{sec:susy}.}
\label{fig:spectra}
\end{figure*}
\begin{figure*}[tbhp]
\centering
\includegraphics[width=\textwidth]{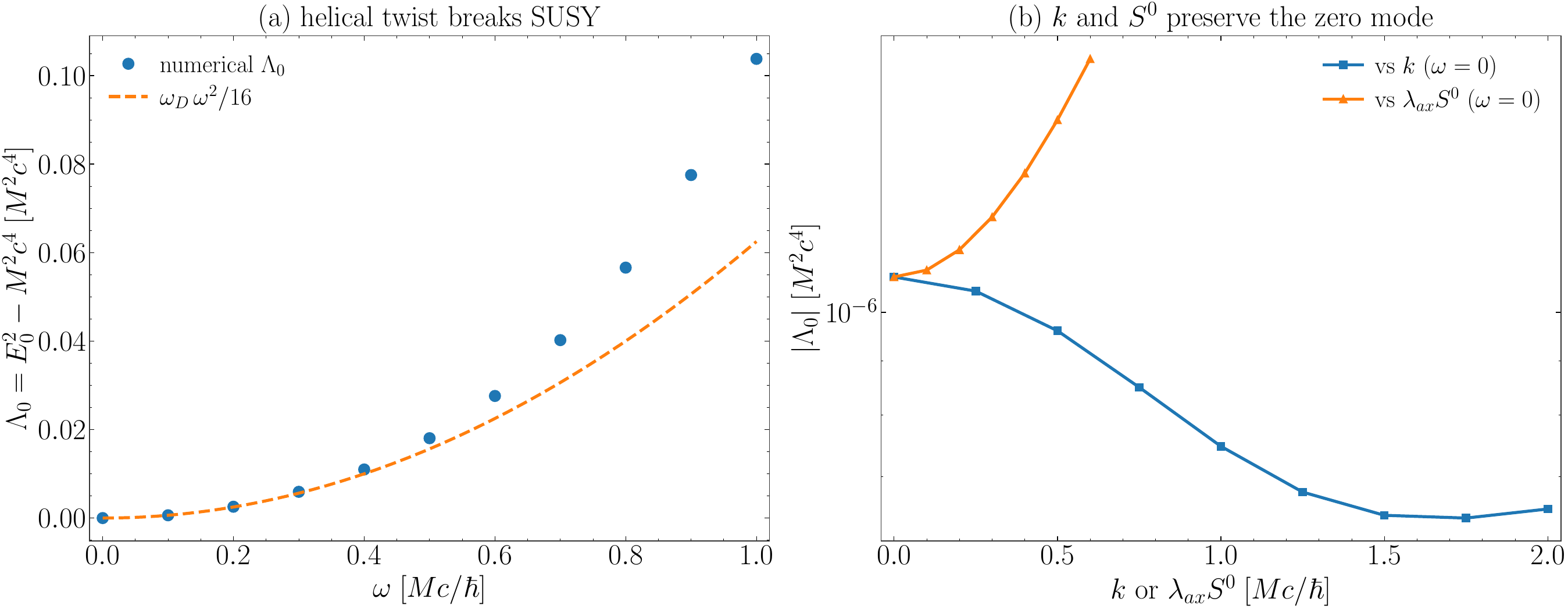}
\caption{Behavior of the supersymmetric zero mode under the three deformations
($\omega_D=1$, $m=1/2$). (a) Ground-state shift $\Lambda_0=E_0^2-M^2c^4$ versus the
helical twist $\omega$ (with $k=S^0=0$): the points follow the quadratic law
$\omega_D\omega^2/16$ (dashed) at small $\omega$, showing that the helical twist
softly breaks supersymmetry. (b) $|\Lambda_0|$ versus the longitudinal momentum $k$
and the axial torsion $\lambda_{\mathrm{ax}}S^0$ (with $\omega=0$): both remain at
the $\sim10^{-6}$ numerical floor, i.e.\ the zero mode, and the exact relation
$E_0=Mc^2$, is preserved.}
\label{fig:zeromode}
\end{figure*}
A subtlety deserves emphasis. Because the radial problem is the coupled spinorial
system~\eqref{eq:coupled-system}, the low-energy eigenstates do not form a single
radial ladder: they organize into two interleaved ladders, one for each spin
sector, distinguished by whether $\psi_{A,1}$ ($\sigma_z=+1$, $\nu_s=0$) or
$\psi_{A,2}$ ($\sigma_z=-1$, $\nu_s=1$) dominates. Table~\ref{tab:states} lists the
six lowest eigenstates ordered by energy. Within each spin sector, the radial node
number grows by one at each step, and the number of maxima of the radial density
equals the number of radial nodes plus one. Since the two ladders interleave, the
peak count is \emph{not} monotonic in the energy ordering: for instance, the third
state ($E=2.387\,Mc^2$) is the nodeless ground state of the $s=-1$ sector and therefore displays a single maximum, even though it lies above two states of the
$s=+1$ sector. In the flat limit, these sector pairs become degenerate [Eq.~\eqref{eq:analytic-flat}], and the helical twist, the longitudinal momentum, and the axial torsion lift the degeneracy. For this reason Fig.~\ref{fig:density}
displays the ladder of a single ($s=+1$) sector, where the expected $n+1$ peak structure is manifest.

\begin{table}[t]
\caption{The six lowest-bound states for the reference parameters
($\omega_D=1$, $\omega=0.3$, $k=0.5$, $m=1/2$, $\lambda_{\mathrm{ax}}S^0=0$),
ordered by energy. The states are split into two spin-sector ladders according to
the dominant spinor component; the number of density maxima equals the number of
radial nodes plus one.}
\label{tab:states}
\centering
\setlength{\tabcolsep}{4pt}
\small
\begin{tabular}{ccccc}
\toprule
state & $E\,[Mc^2]$ & sector & nodes & peaks\\
\midrule
0 & 1.003 & $\psi_{A,1}\ (s{=}{+}1)$ & 0 & 1\\
1 & 2.053 & $\psi_{A,1}\ (s{=}{+}1)$ & 1 & 2\\
2 & 2.387 & $\psi_{A,2}\ (s{=}{-}1)$ & 0 & 1\\
3 & 2.808 & $\psi_{A,1}\ (s{=}{+}1)$ & 2 & 3\\
4 & 3.171 & $\psi_{A,2}\ (s{=}{-}1)$ & 1 & 2\\
5 & 3.411 & $\psi_{A,1}\ (s{=}{+}1)$ & 3 & 4\\
\bottomrule
\end{tabular}
\end{table}

Finally, Fig.~\ref{fig:spectra} summarizes the dependence of the spectrum on the four physically relevant parameters. Panel (a) shows that the helical twist produces relatively mild shifts of the excited states in the interval displayed, while the lowest level is almost pinned to the rest energy. This weak response is consistent with the zero-mode analysis of Sec.~\ref{sec:susy}: the leading lift of the ground state is quadratic in $\omega$, whereas the higher levels are affected both by the Coulomb-like term proportional to $\omega km/r$ and by the spinorial off-diagonal couplings. Panel (b) is the most distinctive. The energy levels are not symmetric under $k\to-k$, and exhibit avoided crossings, which is a direct manifestation of the helical coupling between angular and longitudinal motion through the combination $m/r-\omega k$ and the $\sigma_y$ term
$2M\hbar\omega_Dkr$ in Eq.~\eqref{eq:radial-matrix-final-corrected}. A purely
conical defect would only renormalize the angular quantum number and would not
produce this asymmetric rearrangement of the level spacings.

Panel (c) recovers the expected monotonic growth with the oscillator frequency: increasing $\omega_D$ strengthens the harmonic confinement, raises the excited levels, and enlarges the separation between neighboring radial states. Panel (d) shows that the axial torsion changes the excited levels without destroying the zero-mode structure. Its effect is not a simple rigid shift, because $\lambda_{\rm ax}S^0$ enters both the scalar shift $q(r)$ and the spin-dependent terms in $Z(r)$ and $Y(r)$. Taken together, the four scans demonstrate that the
longitudinal momentum is the most efficient probe of the helical geometry, while $\omega_D$ controls the global confinement scale, and the axial torsion produces a more moderate spinorial splitting.

\section{Supersymmetric structure and the helical zero mode}
\label{sec:susy}

The first-order system~\eqref{eq:first-order-corrected} possesses the algebraic
structure of supersymmetric quantum mechanics. The operators $\hat\Pi_+$ and
$\hat\Pi_-$ intertwine the upper and lower spinors, and the second-order operators
of Eq.~\eqref{eq:second-order-corrected-master} are the partner Hamiltonians
\begin{equation}
\hat H_A=\hat\Pi_+\hat\Pi_-,\qquad \hat H_B=\hat\Pi_-\hat\Pi_+,
\label{eq:partners}
\end{equation}
which share the same nonnegative spectrum $\Lambda=E^2-M^2c^4\ge0$, except possibly for a zero mode. A state with $E=Mc^2$ (i.e.\ $\Lambda=0$) must be annihilated by the charge operator, $\hat\Pi_-\psi_A=0$ with $\psi_B=0$; supersymmetry is unbroken when such a normalizable zero mode exists, and its presence pins the lowest level exactly at the rest energy.
\begin{figure*}[tbhp]
\centering
\includegraphics[width=0.48\textwidth]{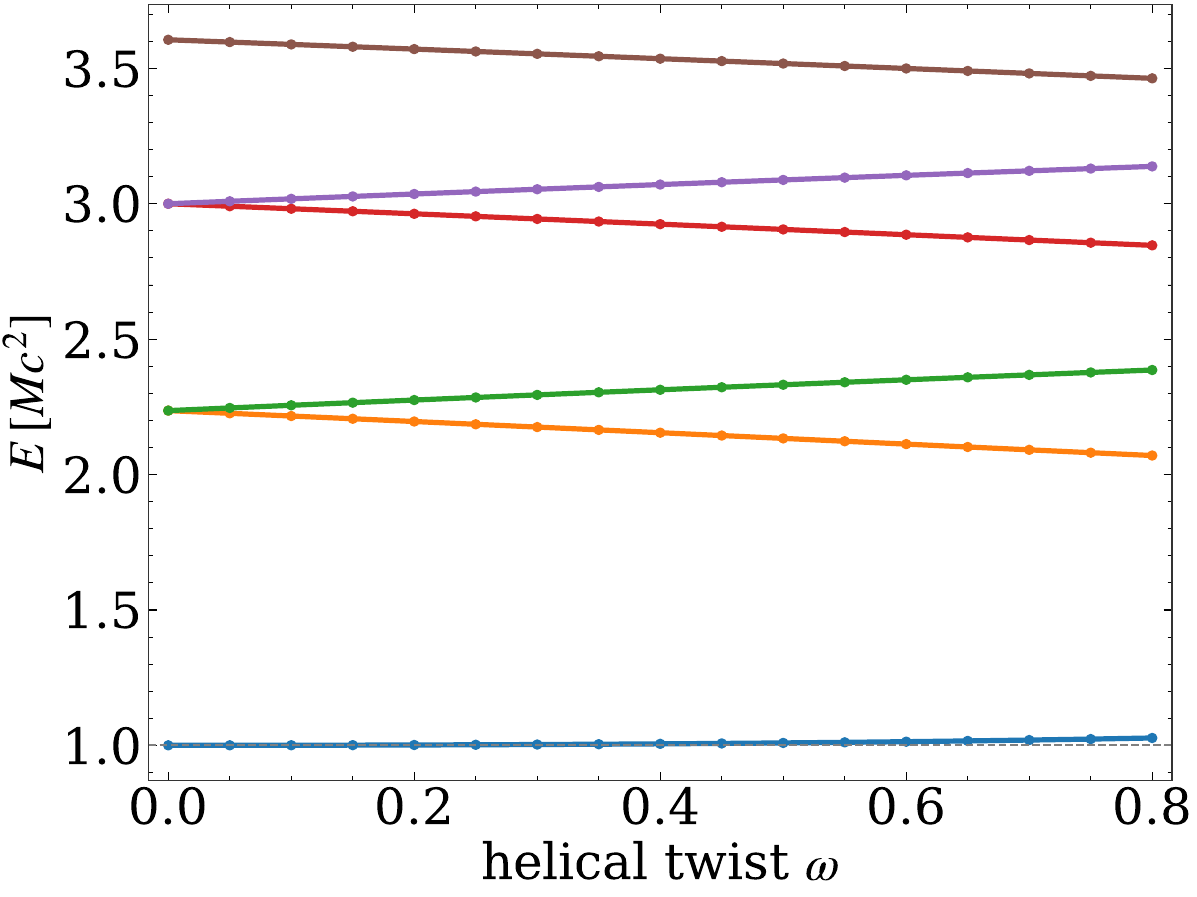}\hfill
\includegraphics[width=0.48\textwidth]{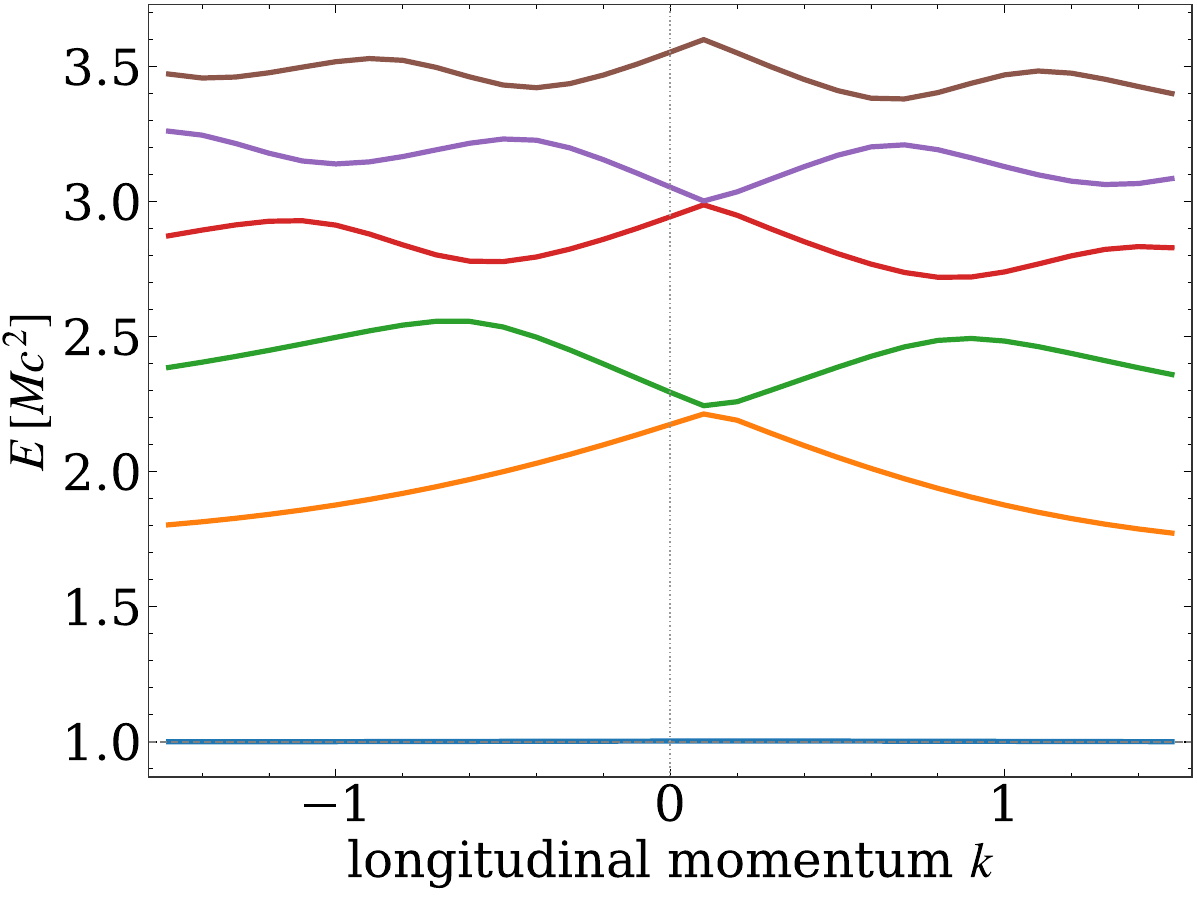}
\caption{Spectral flow of the six lowest bound-state energies (in units of $Mc^2$)
for the reference parameters $\omega_D=1$, $m=1/2$, $\lambda_{\mathrm{ax}}S^0=0$.
(a) Flow versus the helical twist $\omega$ at $k=0$: the ground state stays at
$E\simeq Mc^2$ (dashed line), confirming the soft, quadratic supersymmetry
breaking, while the excited levels split and disperse. (b) Flow versus the
longitudinal momentum $k$ at $\omega=0.3$: the levels are asymmetric under
$k\to-k$ and develop avoided crossings, the spectral fingerprint of the helical
combination $m/r-\omega k$. The two panels are the continuous counterpart of the
parameter scans in Fig.~\ref{fig:spectra}(a,b).}
\label{fig:flow}
\end{figure*}

In the flat limit ($\omega=0$, $S^0=0$, $k=0$) the system decouples into the two
spin sectors and reduces to the planar Dirac oscillator. Writing $u=\sqrt r\,\psi$,
each sector factorizes as $\hat H=\mathcal A^{\dagger}\mathcal A$ with
\begin{equation}
\mathcal A=\hbar c\left(\frac{d}{dr}+W\right),\qquad
W(r)=\frac{M\omega_D}{\hbar}\,r-\frac{\nu_s+\tfrac12}{r}.
\label{eq:superpotential}
\end{equation}
The zero mode $\mathcal A\,u_0=0$ reads
$u_0\propto r^{\nu_s+1/2}e^{-M\omega_D r^2/2\hbar}$, i.e.
$\psi_0\propto r^{\nu_s}e^{-M\omega_D r^2/2\hbar}$, which is normalizable for the
flat spin sector $\nu_s=0$ ($s=+1$, $l=0$). We use the word ``sector'' for the
eigenspaces that become definite $\sigma_z$ sectors in the flat limit; away from
that limit the components are coupled, but the dominant component still provides a
useful label for the numerical states. This is the exact supersymmetric ground state,
with $E_0=Mc^2$, consistent with the spectrum~\eqref{eq:analytic-flat} and with the
nodeless density of Fig.~\ref{fig:density}.

Whether this zero mode survives the helical and torsional deformations is answered
unambiguously by solving the coupled system~\eqref{eq:coupled-system}; the result is
summarized in Fig.~\ref{fig:zeromode}. The longitudinal momentum $k$ and the axial
torsion $\lambda_{\mathrm{ax}}S^0$ leave the ground-state shift at the level of the
discretization floor, $|\Lambda_0|\lesssim10^{-6}$: the supersymmetric zero mode
persists, and $E_0=Mc^2$ remains exact. The helical twist $\omega$, by contrast,
lifts the zero mode, and the lift grows quadratically. In natural units
($\hbar=c=M=1$) we find
\begin{equation}
\Lambda_0=E_0^2-M^2c^4\simeq\frac{1}{16}\,\omega_D\,\omega^2+\mathcal O(\omega^4),
\label{eq:zeromode-lift}
\end{equation}
so that $E_0$ departs from the rest energy only through the geometry.
The mechanism can be understood directly from the first-order equation for the
zero mode. For $k=0$ the axial torsion contributes only a constant scalar shift to
$q(r)$, whereas the helical twist contributes the singular profile
$\hbar c\omega/4r$. A constant shift can be absorbed in the first-order zero-mode
equation without changing normalizability, but the $1/r$ profile modifies the
small-$r$ behavior of the solution and destroys the exact kernel of
$\hat\Pi_-$. Among the three control parameters, the helical (screw-dislocation) twist is
thus the unique supersymmetry-breaking agent, and it breaks it softly, as
$\omega^2$. This also refines Fig.~\ref{fig:spectra}: the lowest level is exactly
pinned at $E=Mc^2$ along the $k$ and $S^0$ scans, and only weakly displaced
($\propto\omega^2$) along the $\omega$ scan.

The continuous spectral flow of the six lowest levels is displayed in
Fig.~\ref{fig:flow}. Panel~(a) tracks the levels as the helical twist $\omega$ is
switched on at $k=0$: the ground state stays pinned at $E\simeq Mc^2$ (dashed line)
up to the soft quadratic shift discussed above, while the excited doublets split
and disperse as the angular--longitudinal mixing lifts the flat-limit degeneracy.
Panel~(b) follows the same levels along the longitudinal momentum $k$ at
$\omega=0.3$: the spectrum is manifestly asymmetric under $k\to-k$, the extrema of
each branch are displaced from $k=0$, and neighboring branches form avoided
crossings rather than touching. This continuous picture is the counterpart of the
discrete parameter scans in Fig.~\ref{fig:spectra}(a,b) and makes explicit that the
helical twist is the only deformation that moves the protected ground state.

\section{Thermodynamic properties}
\label{sec:thermo}

The discrete spectrum obtained from the radial problem also allows one to define thermal observables. Since the present calculation is performed at fixed azimuthal and longitudinal quantum numbers, the thermodynamic quantities below are sector-resolved: they characterize a single $(m,k)$ sector rather than the full many-sector gas. For each fixed sector, we use the positive-energy levels $\{E_n(m,k)\}$ as a one-particle spectrum coupled to a reservoir at temperature $T$. With energies measured in units of $Mc^2$, the dimensionless temperature and partition functions are
\begin{align}
\tau &\equiv \frac{k_BT}{Mc^2},
\label{eq:tau-def}\\
Z_{m,k}(\tau)&=\sum_{n=0}^{N_{\rm lev}-1}
\exp\left[-\frac{E_n(m,k)}{\tau}\right].
\label{eq:sector-partition}
\end{align}
In the numerical evaluation, we take $N_{\rm lev}=50$, which is sufficient for the
range $\tau\lesssim1.2$ shown below. The normalized Boltzmann weights are
\begin{equation}
p_n(\tau)=\frac{\exp[-E_n(m,k)/\tau]}{Z_{m,k}(\tau)}.
\label{eq:boltzmann-weights}
\end{equation}
The sector-resolved Helmholtz free energy, internal energy, entropy, and heat capacity can then be written in the numerically stable form
\begin{align}
F_{m,k} &= -\tau\ln Z_{m,k},
\label{eq:F-sector}\\
U_{m,k} &= \sum_n p_n E_n
        = \tau^2\frac{\partial\ln Z_{m,k}}{\partial\tau},
\label{eq:U-sector}\\
S_{m,k} &= \ln Z_{m,k}+\frac{U_{m,k}}{\tau},
\label{eq:S-sector}\\
C_{m,k} &= \frac{1}{\tau^2}
\left(\sum_n p_nE_n^2-U_{m,k}^2\right).
\label{eq:C-sector}
\end{align}
Equations~\eqref{eq:S-sector} and~\eqref{eq:C-sector} are especially useful for
interpreting the numerical curves: the entropy measures the number of thermally
accessible states in the chosen sector, while the heat capacity is controlled by
the variance of the occupied energy levels. Thus $C_{m,k}$ is most sensitive to
changes in the first few gaps rather than to a uniform shift of the whole
spectrum.

\begin{figure*}[tbhp]
\centering
\includegraphics[width=\textwidth]{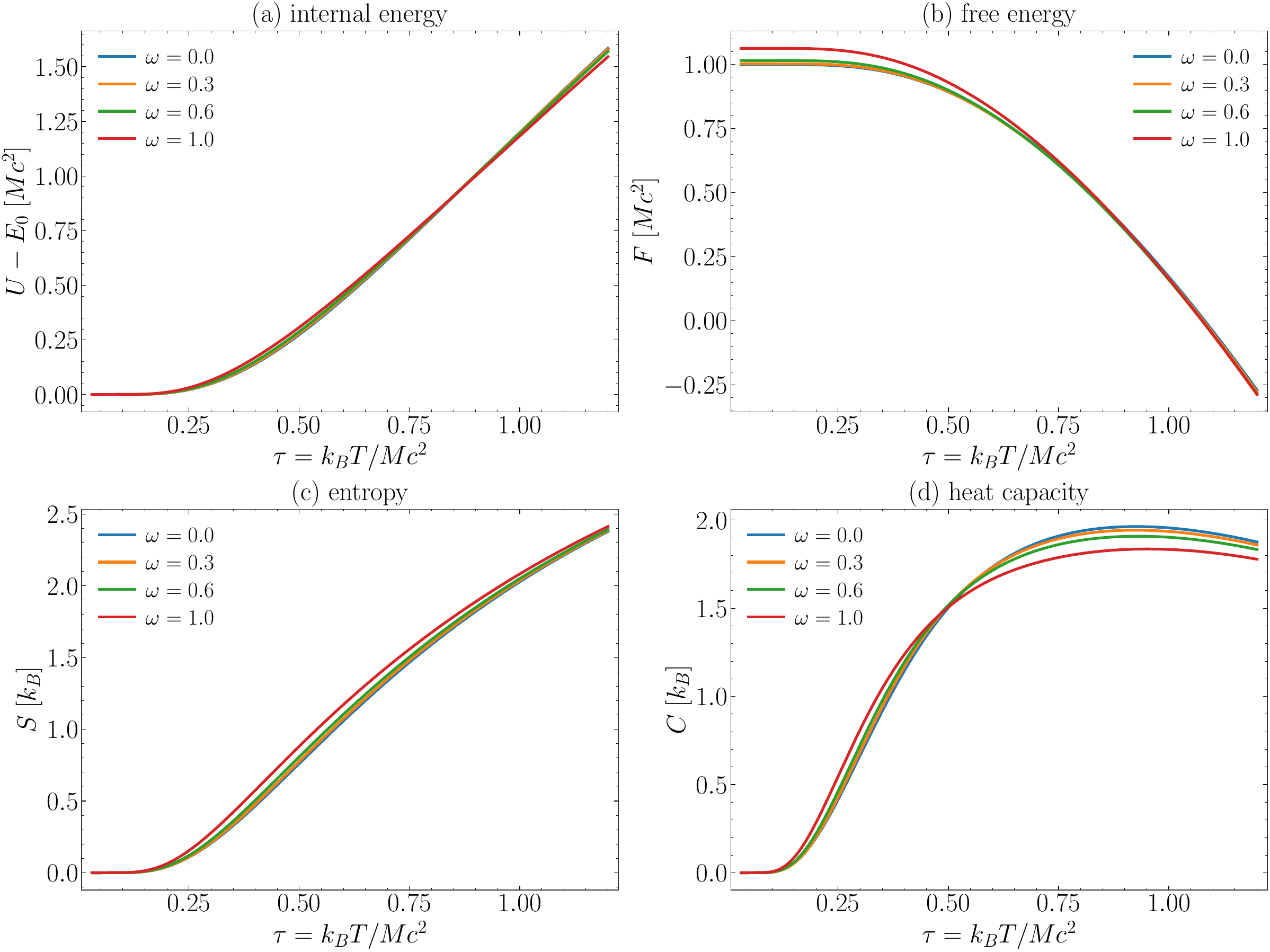}
\caption{Sector-resolved thermodynamic functions versus the dimensionless temperature $\tau=k_BT/Mc^2$ for several values of the helical twist $\omega$ ($\omega_D=1$, $k=0.5$, $m=1/2$, $\lambda_{\rm ax}S^0=0$). (a) Internal energy measured from the ground state, $U_{m,k}-E_0$; (b) Helmholtz free energy $F_{m,k}$; (c) entropy $S_{m,k}$; and (d) heat capacity $C_{m,k}$. The low-temperature plateau reflects the isolated ground state, whereas the broad maximum in $C_{m,k}$ signals thermal activation of the first few excited levels.}
\label{fig:thermo}
\end{figure*}

\begin{figure*}[tbhp]
\centering
\includegraphics[width=\textwidth]{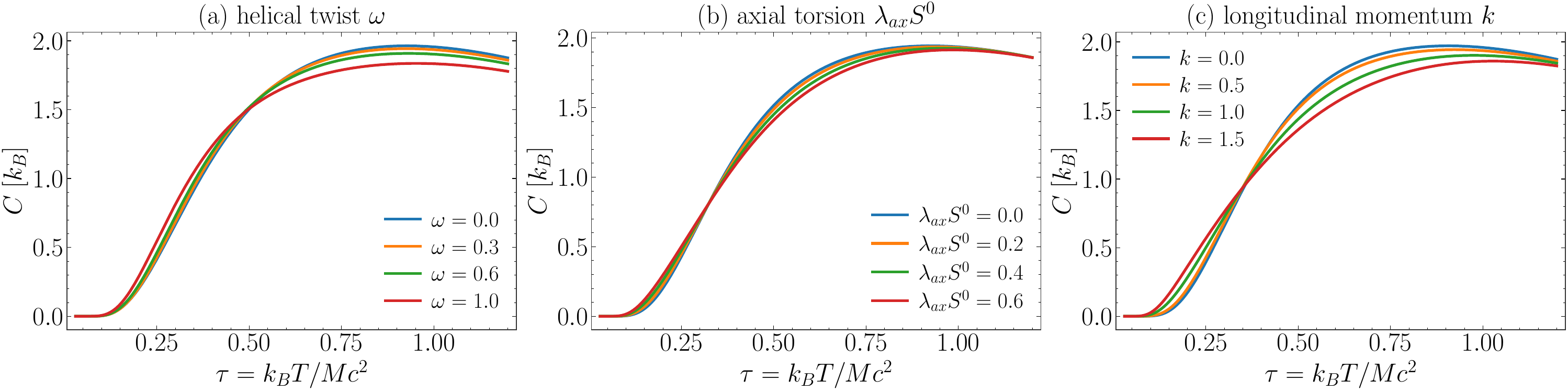}
\caption{Heat capacity $C_{m,k}(\tau)$ showing the separate thermal signatures of
the three control parameters (reference $\omega_D=1$, $\omega=0.3$, $k=0.5$,
$m=1/2$, $\lambda_{\rm ax}S^0=0$, varying one at a time): (a) helical twist
$\omega$, (b) axial torsion $\lambda_{\rm ax}S^0$, and (c) longitudinal momentum
$k$. The helical twist and the axial torsion produce moderate peak shifts,
whereas $k$ significantly changes the peak height and width because it strongly
reorganizes the low-energy spacings through the helical combination
$m/r-\omega k$.}
\label{fig:heatcap}
\end{figure*}

Figure~\ref{fig:thermo} displays the thermal functions obtained by varying the
helical twist while keeping $\omega_D=1$, $k=0.5$, $m=1/2$, and
$\lambda_{\rm ax}S^0=0$. At low temperature, only the ground state contributes:
$U_{m,k}-E_0\to0$, $S_{m,k}\to0$, and $C_{m,k}\to0$, while the free energy tends to
the ground-state energy. As $\tau$ increases, the first excited levels become
thermally populated and $U_{m,k}$ grows smoothly. The free energy decreases
because the entropic contribution $-\tau S_{m,k}$ becomes progressively more
important. The entropy rises rapidly once several low-lying levels participate in
the ensemble and then increases more slowly over the plotted range, reflecting
the finite number of states effectively sampled for $\tau\lesssim1.2$.

The heat capacity in Fig.~\ref{fig:thermo}(d) exhibits a broad Schottky-type
maximum. Physically, this maximum occurs when the thermal scale is comparable to
the dominant gaps between the ground state and the first few excited states. The
peak is broad rather than sharp because the spectrum contains several nearby
level spacings instead of a single two-level gap. Increasing the helical twist
slightly lowers the maximum and shifts the curve, but the overall effect is mild
in this scan. This is consistent with Fig.~\ref{fig:spectra}(a): for the chosen
parameters, $\omega$ weakly modifies the low-energy gaps, while the zero-mode
sector remains nearly pinned to $E=Mc^2$ up to the soft quadratic lift discussed
in Sec.~\ref{sec:susy}.

Figure~\ref{fig:heatcap} compares the heat-capacity response to each control
parameter separately. The helical twist, panel (a), mainly produces a modest
reduction of the high-temperature side of the peak, indicating a small
redistribution of the first few level spacings. The axial torsion, panel (b), has
an even smoother effect for the range considered: it changes the spinorial
splitting through the constant axial shift but does not qualitatively alter the
thermal activation pattern. The longitudinal momentum, panel (c), produces the
largest change. Increasing $k$ shifts the onset of thermal excitation and broadens
the peak because $k$ enters the radial Hamiltonian both through the helical
combination $m/r-\omega k$ and through the off-diagonal term proportional to
$kr\,\sigma_y$. The same mechanism is responsible for the strong spectral
asymmetry and avoided crossings in Fig.~\ref{fig:spectra}(b). Hence, the thermal
curves provide an independent diagnostic of the angular--longitudinal mixing.

These results should be interpreted as sector-resolved thermodynamics. A full canonical treatment of an extended system would require summing over all allowed angular sectors and replacing the fixed longitudinal wave number by either a finite-size quantization along $z$ or a density-of-states integral over $k$. The fixed-sector analysis is nevertheless useful because it isolates the effect of each geometric or torsional parameter on the low-energy relativistic spectrum, which is precisely the part of the spectrum controlling the low- and
intermediate-temperature response.

\section{Longitudinal vector and axial currents}
\label{sec:currents}

The numerical eigenfunctions contain more information than the energy spectrum. They also determine the local vector and axial currents carried by each confined state. In a twisted background with a distinguished longitudinal direction, the most relevant components are
\begin{equation}
 j^z=\bar\Psi\gamma^z\Psi,
 \qquad
 j_5^z=\bar\Psi\gamma^z\gamma^5\Psi,
\label{eq:current-defs}
\end{equation}
where $\bar\Psi=\Psi^\dagger\gamma^0$. The metric determinant is
$g\equiv\det(g_{\mu\nu})=-c^2r^2$, so that $\sqrt{-g}=cr$ and the same radial
measure $r\,dr$ used in the normalization enters the integrated currents per
unit length along the axis.

Using the curved gamma matrix $\gamma^z=-\omega\gamma^2+\gamma^3$, one finds that the longitudinal current operators are governed by the single Pauli-block matrix
\begin{equation}
T\equiv \sigma_z-\omega\sigma_y,
\label{eq:T-current}
\end{equation}
through
\begin{equation}
\gamma^0\gamma^z=\begin{pmatrix}0&T\\ T&0\end{pmatrix},
\qquad
\gamma^0\gamma^z\gamma^5=\begin{pmatrix}T&0\\0&T\end{pmatrix}.
\label{eq:current-blocks}
\end{equation}
For a separated spinor $\Psi=(\psi_A,\psi_B)^{\mathsf T}$, the local densities
therefore become
\begin{align}
 j^z_n(r)&=2\,\mathrm{Re}\!\left[\psi_{A,n}^{\dagger}T\psi_{B,n}\right],
\label{eq:jz-density}\\
 j_{5,n}^z(r)&=\psi_{A,n}^{\dagger}T\psi_{A,n}
 +\psi_{B,n}^{\dagger}T\psi_{B,n},
\label{eq:j5z-density}
\end{align}
where the lower component is reconstructed from the first-order equation,
\begin{equation}
\psi_{B,n}=\frac{1}{E_n+Mc^2}\hat\Pi_-\psi_{A,n}.
\label{eq:psiB-reconstruct}
\end{equation}
The corresponding integrated currents, with the common angular normalization
suppressed, are
\begin{equation}
 \mathcal I_n^z=\int_0^\infty j_n^z(r)\,r\,dr,
 \qquad
 \mathcal I_{5,n}^z=\int_0^\infty j_{5,n}^z(r)\,r\,dr.
\label{eq:integrated-currents}
\end{equation}
For a thermal population in a fixed $(m,k)$ sector, the sector-resolved currents
are obtained with the same Boltzmann weights used in Sec. \ref{sec:thermo},
\begin{equation}
 \langle\mathcal I^z\rangle_{m,k}=\sum_n p_n\mathcal I_n^z,
 \qquad
 \langle\mathcal I_5^z\rangle_{m,k}=\sum_n p_n\mathcal I_{5,n}^z.
\label{eq:thermal-currents}
\end{equation}

These formulas give a direct physical interpretation to the spectral results. First, an exact supersymmetric zero mode with $E=Mc^2$ has $\hat\Pi_-\psi_A=0$ and therefore $\psi_B=0$. It follows immediately from Eq.~\eqref{eq:jz-density} that this state carries no longitudinal vector current, $\mathcal I_0^z=0$. Its axial current, however, is generally nonzero. In the flat $s=+1$ sector, where $T\to\sigma_z$ and $\psi_A\propto(1,0)^{\mathsf T}\exp[-M\omega_Dr^2/(2\hbar)]$, the normalized zero mode gives $\mathcal I_{5,0}^z=1$. The supersymmetric ground state is thus axially polarized even though it carries no charge current along the axis.

\begin{figure*}[t!]
\centering
\includegraphics[width=0.48\textwidth]{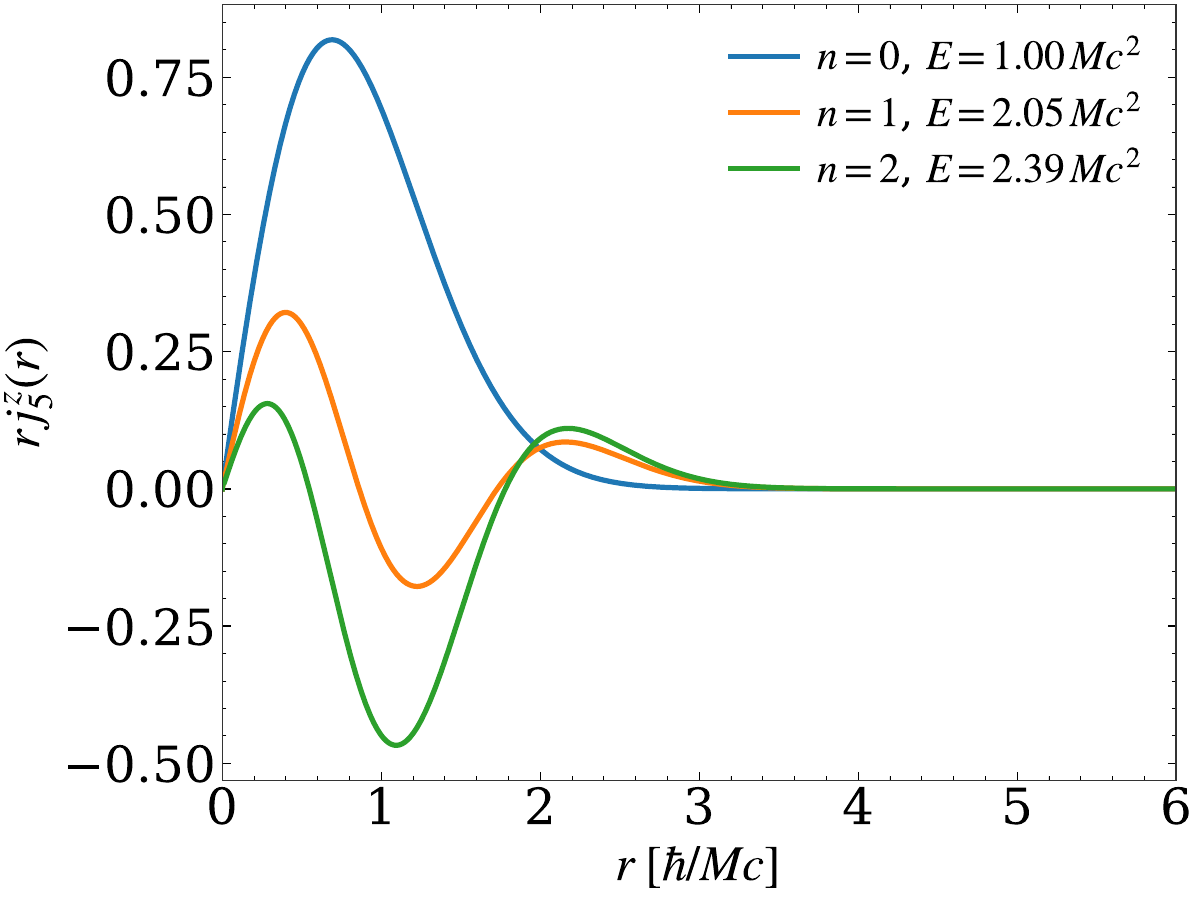}\hfill
\includegraphics[width=0.48\textwidth]{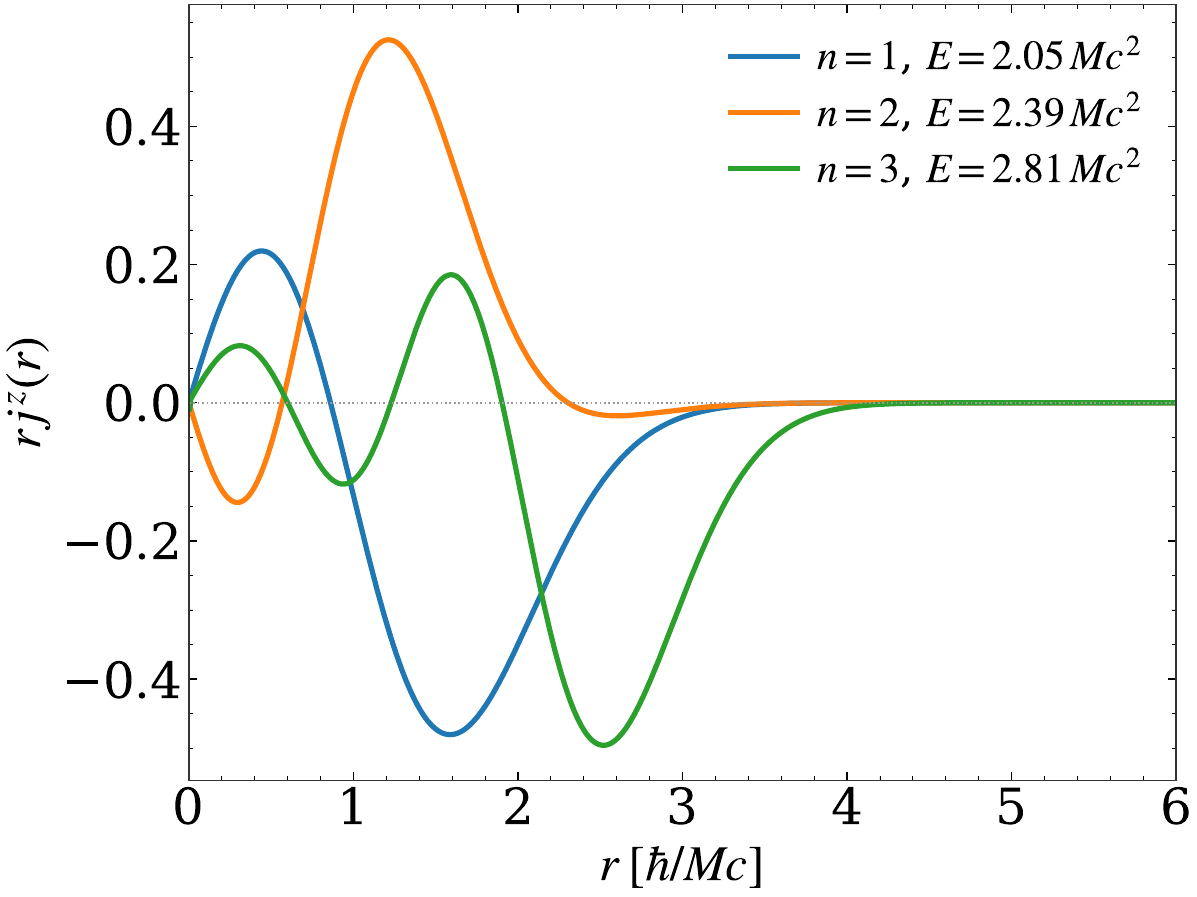}\\[6pt]
\includegraphics[width=0.48\textwidth]{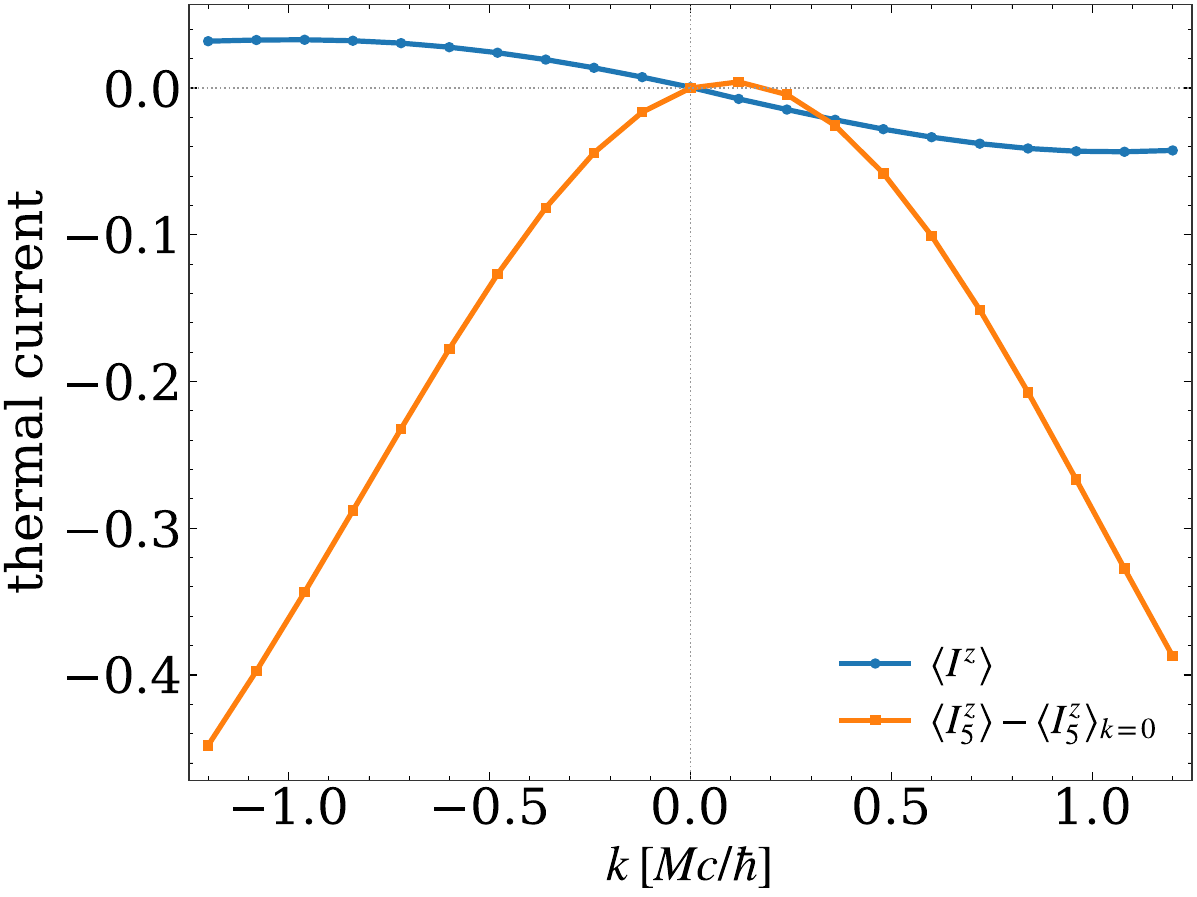}\hfill
\includegraphics[width=0.48\textwidth]{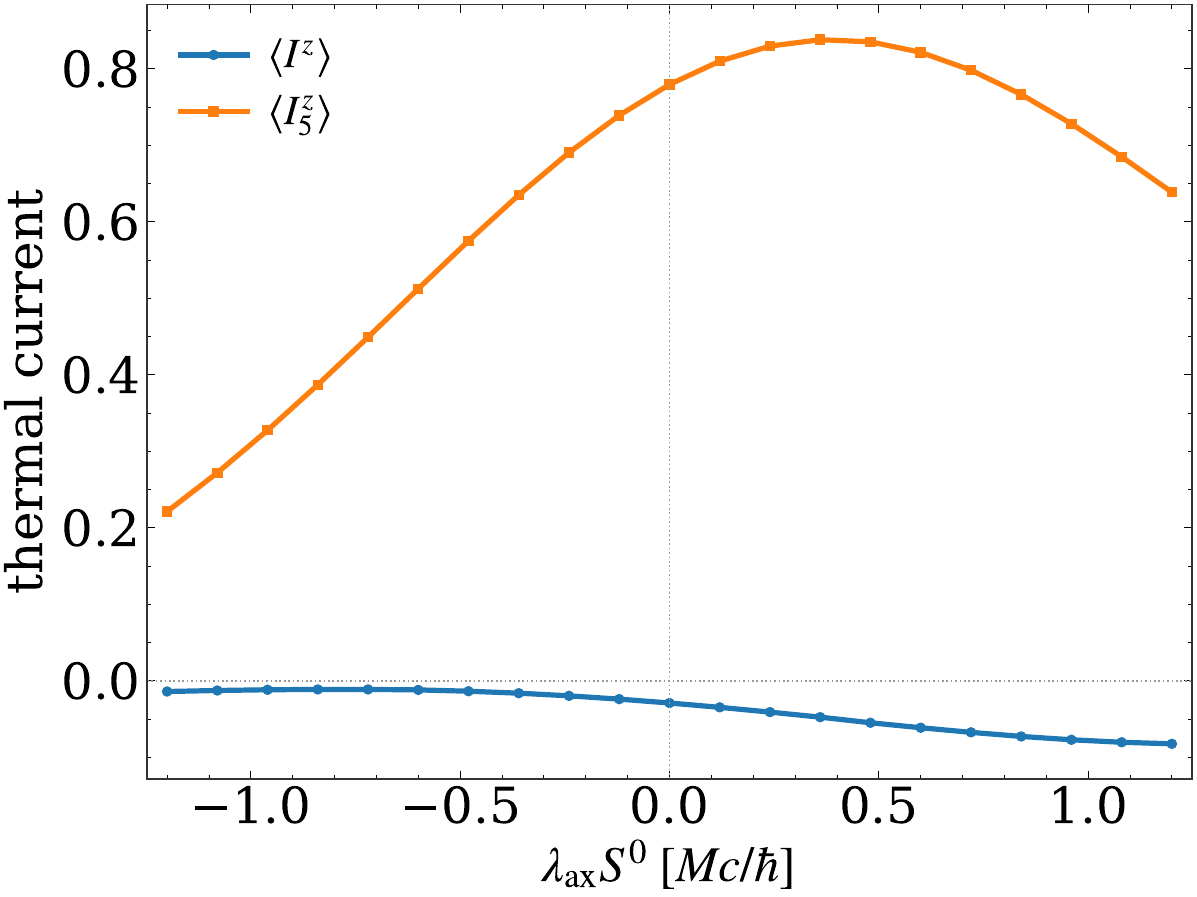}
\caption{Longitudinal vector and axial currents of the bound states, computed from
the reconstructed spinors via Eqs.~\eqref{eq:jz-density}--\eqref{eq:thermal-currents}
at the reference parameters $\omega_D=1$, $\omega=0.3$, $k=0.5$, $m=1/2$,
$\lambda_{\mathrm{ax}}S^0=0$ (except for the parameter being scanned). (a) Axial
current density $r\,j_5^z(r)$ for the lowest states $n=0,1,2$; the nodeless $n=0$
curve is the axially polarized supersymmetric zero mode with $\mathcal I_{5,0}^z=1$.
(b) Vector (charge) current density $r\,j^z(r)$ for $n=1,2,3$, which integrates to a
finite longitudinal current for the excited states while vanishing for the zero
mode. (c) Thermally averaged currents $\langle\mathcal I^z\rangle$ and
$\langle\mathcal I_5^z\rangle$ versus the longitudinal momentum $k$ (the axial
curve is shown relative to its $k=0$ value); the vector current is odd in $k$, the
transport fingerprint of the helical $k\to-k$ asymmetry. (d) Thermally averaged
currents versus the axial torsion $\lambda_{\mathrm{ax}}S^0$, showing the growth of
the axial polarization with torsion and the comparatively weak vector response.}
\label{fig:currents}
\end{figure*}

Second, the helical twist converts the longitudinal momentum into a genuine
transport probe. Since the spectrum in Fig.~\ref{fig:spectra}(b) is not invariant
under $k\to -k$, the current in a fixed $(m,k)$ sector is not generally canceled
by a partner state at the same energy. The origin of this asymmetry is precisely
the combination $m/r-\omega k$ in $P_\phi$ and the off-diagonal
$kr\,\sigma_y$ coupling in the radial Hamiltonian. A full ensemble symmetric in
$k$ would cancel the odd part of the vector current, but a sector-resolved or
longitudinally biased population retains a net axial-direction response; this
asymmetry is shown quantitatively in Fig.~\ref{fig:currents}(c).

Finally, the axial torsion controls the chiral sector directly because it couples
to $\bar\Psi\gamma^\mu\gamma^5\Psi$. In the present bound-state problem, this does
not amount to computing an anomaly coefficient, since no Dirac sea regularization is performed. Nevertheless, Eqs.~\eqref{eq:j5z-density} and \eqref{eq:thermal-currents} provide finite observables that quantify the axial polarization induced by $\lambda_{\rm ax}S^0$. This is the confined-spectrum counterpart of the axial--torsional responses discussed in Weyl and Dirac materials with effective torsion~\cite{PRL.2019.122.056601,PRR.2020.2.022016,PRD.2025.112.081903}; its dependence on $\lambda_{\rm ax}S^0$ is displayed in Fig.~\ref{fig:currents}(d).

The local current densities and their integrated, thermally averaged responses are collected in Fig.~\ref{fig:currents}. Panels~(a) and~(b) show the axial and vector current densities $r\,j_5^z(r)$ and $r\,j^z(r)$ for the lowest bound states at the reference parameters. The axial density of the supersymmetric ground state [$n=0$ in panel~(a)] is nodeless and integrates to $\mathcal I_{5,0}^z=1$, while its vector current vanishes identically, consistent with $\psi_B=0$; the excited states develop additional radial nodes that mirror the density ladder of Fig.~\ref{fig:density}. Panel~(c) displays the thermal currents $\langle\mathcal I^z\rangle$ and $\langle\mathcal I_5^z\rangle$ as functions of the longitudinal momentum $k$: the vector current is manifestly odd and does not cancel within a fixed $(m,k)$ sector, the quantitative signature of the $k\to-k$ asymmetry. Panel~(d) shows the monotonic build-up of the axial response with the axial torsion $\lambda_{\rm ax}S^0$, together with a much weaker vector current, confirming that torsion polarizes the chiral sector while leaving the longitudinal charge transport nearly unaffected.

\section{Witten index and zero-mode protection}
\label{sec:index}

The supersymmetric structure of Sec.~\ref{sec:susy} can be expressed more
sharply in terms of the Witten index~\cite{NPB.1981.188.3.513,PhysRep.1995.251.5-6.267}. We introduce
the supercharges
\begin{equation}
\mathcal Q=\begin{pmatrix}0&0\\ \hat\Pi_-&0\end{pmatrix},
\qquad
\mathcal Q^\dagger=\begin{pmatrix}0&\hat\Pi_+\\0&0\end{pmatrix},
\label{eq:supercharges}
\end{equation}
so that
\begin{equation}
\{\mathcal Q,\mathcal Q^\dagger\}=\begin{pmatrix}
\hat H_A&0\\0&\hat H_B
\end{pmatrix},
\qquad
\mathcal Q^2=(\mathcal Q^\dagger)^2=0,
\label{eq:susy-algebra}
\end{equation}
with $\hat H_A=\hat\Pi_+\hat\Pi_-$ and
$\hat H_B=\hat\Pi_-\hat\Pi_+$. The Witten index is
\begin{equation}
\Delta=\dim\ker\hat H_A-\dim\ker\hat H_B.
\label{eq:witten-index}
\end{equation}
In this radial problem, the index should be understood as a spectral diagnostic of the supersymmetric partner operators, not as a statement about the global topology of spacetime. It counts the unpaired zero modes of the factorized radial Hamiltonians are stable under deformations that preserve the operator domain and the normalizability of the zero-mode kernel.

In the flat Dirac-oscillator limit, $\omega=S^0=k=0$, the sector $\nu_s=0$ contains one normalizable solution of $\hat\Pi_-\psi_A=0$, while the partner equation has no normalizable zero mode in the corresponding sector. Thus
\begin{equation}
\Delta=1
\end{equation}
for the supersymmetric ground-state sector. Positive-$\Lambda$ states are paired
between $\hat H_A$ and $\hat H_B$, whereas the zero mode remains unpaired and pins
the lowest relativistic energy at $E=Mc^2$.

The numerical results in Fig.~\ref{fig:zeromode} show how the three deformations
affect this index structure. The longitudinal momentum $k$ and the constant axial
torsion shift $\lambda_{\rm ax}S^0$ preserve the zero-mode kernel: within numerical
accuracy, $|\Lambda_0|\lesssim10^{-6}$ and the unpaired state remains at
$E=Mc^2$. In this sense, they preserve the index of the radial supersymmetric
sector. The helical twist is different. It contributes to the nonconstant profile
$\hbar c\omega/(4r)$ to $q(r)$, changes the small-$r$ behavior of the first-order
zero-mode equation, and removes the exact normalizable kernel of $\hat\Pi_-$. The index of the deformed sector then becomes effectively trivial, and the ground state is lifted according to Eq.~\eqref{eq:zeromode-lift}. Thus, the Witten-index viewpoint gives a compact explanation of why axial torsion and longitudinal momentum preserve the supersymmetric zero mode, whereas the helical twist is the unique supersymmetry-breaking deformation in the present model.

\section{Nonrelativistic limit and emergent geometric couplings}
\label{sec:nrlimit}

Although the main formulation developed above is fully relativistic, it is useful
to display the leading nonrelativistic limit in order to identify the effective
geometric and torsional structures contained in the radial Hamiltonian. We write
\begin{equation}
E=Mc^2+\varepsilon,\qquad \varepsilon\ll Mc^2,
\end{equation}
so that
\begin{equation}
\Lambda=E^2-M^2c^4=2Mc^2\varepsilon+\mathcal O(\varepsilon^2).
\end{equation}
Dividing the coupled radial system~\eqref{eq:coupled-system} by $2Mc^2$, and
keeping the leading terms in $\varepsilon/Mc^2$, gives a Schr\"odinger-type
equation for the upper Pauli spinor,
\begin{equation}
\hat H_{\rm NR}\psi_A=\varepsilon\psi_A.
\label{eq:HNR-eigen}
\end{equation}
For compactness, let
\begin{equation}
\chi\equiv \lambda_{\rm ax}S^0 .
\end{equation}
The effective Hamiltonian can then be written as
\begin{equation}
\hat H_{\rm NR}=\hat H_{\rm sc}\,\mathbb I_2
+\hat H_x\sigma_x+\hat H_y\sigma_y+\hat H_z\sigma_z,
\label{eq:HNR-decomp}
\end{equation}
where the scalar part is
\begin{align}
\hat H_{\rm sc}={}&-
\frac{\hbar^2}{2M}\left(\frac{d^2}{dr^2}
+\frac{1}{r}\frac{d}{dr}\right)
+\frac{\hbar^2}{2Mr^2}\left(m^2+\frac14+\frac{\omega^2}{16}\right)
\nonumber\\
&-\frac{\hbar^2\omega(4km+\chi)}{4Mr}
+\frac12M\omega_D^2r^2-\frac{\hbar\omega_D}{2}
\nonumber\\
&+\frac{\hbar^2(1+\omega^2)k^2}{2M}
+\frac{\hbar^2\chi^2}{2M}.
\label{eq:HNR-scalar}
\end{align}
The spin-dependent terms are
\begin{align}
\hat H_x={}&-
\frac{i\hbar^2}{2M}
\left[\left(\frac{\omega}{2r}-2\chi\right)\frac{d}{dr}
-\frac{\chi}{r}\right],
\label{eq:HNR-x}
\\
\hat H_y={}&
\frac{\hbar^2m\omega}{4Mr^2}
-\frac{\hbar^2k\omega^2}{4Mr}
-\frac{\hbar^2m\chi}{Mr}
+\hbar\omega_Dkr
+\frac{\hbar^2k\omega\chi}{M},
\label{eq:HNR-y}
\\
\hat H_z={}&-
\frac{\hbar^2m}{2Mr^2}
+\frac{\hbar^2k\omega}{4Mr}
+\hbar\omega_D\omega kr
-\hbar\omega_Dm
-\frac{\hbar^2k\chi}{M}.
\label{eq:HNR-z}
\end{align}
Equations~\eqref{eq:HNR-decomp}--\eqref{eq:HNR-z} make explicit the terms that
were hidden in the compact relativistic radial equation. The off-diagonal pieces
$\hat H_x$ and $\hat H_y$ are retained explicitly; they originate from the
$\sigma_x$ derivative coupling and from the $\sigma_y$ mixing term in
Eq.~\eqref{eq:coupled-system}. Therefore, the nonrelativistic limit remains, in
general, a two-component problem whenever the helical twist, axial torsion, or
longitudinal momentum couples the two spin components.

The scalar part contains an emergent Coulomb-like contribution,
\begin{equation}
V_C(r)=-\frac{\hbar^2\omega(4km+\chi)}{4Mr},
\label{eq:HNR-coulomb}
\end{equation}
which is the nonrelativistic remnant of the geometric term generated by the
off-diagonal metric and of its axial-torsion partner. The centrifugal barrier is
also renormalized by the helical spin-connection contribution $\omega^2/16$, while $\hat H_z$ contains the familiar Dirac-oscillator spin-orbit term $-\hbar\omega_Dm\sigma_z$ together with the geometric linear coupling $\hbar\omega_D\omega kr\,\sigma_z$. These terms show how the helical twist reorganizes the low-energy dynamics: it not only mixes angular and
longitudinal motion through $m/r-\omega k$, but also induces Coulomb-like and
spin-dependent potentials in the Schr\"odinger regime.

As a consistency check, setting $\omega=k=\chi=0$ diagonalizes
Eq.~\eqref{eq:HNR-decomp} in the $\sigma_z$ basis. For a spin sector
$s=\pm1$, one obtains the planar nonrelativistic Dirac-oscillator limit,
\begin{align}
\hat H^{(s)}_{\rm NR}={}&-
\frac{\hbar^2}{2M}\left(\frac{d^2}{dr^2}
+\frac{1}{r}\frac{d}{dr}\right)
+\frac{\hbar^2}{2Mr^2}\left(m-\frac{s}{2}\right)^2
\nonumber\\
&+\frac12M\omega_D^2r^2
-\hbar\omega_D\left(\frac12+sm\right),
\label{eq:HNR-flat}
\end{align}
whose spectrum follows directly from the nonrelativistic expansion of
Eq.~\eqref{eq:analytic-flat},
\begin{equation}
\varepsilon_{n,s}=\hbar\omega_D\bigl(2n+|\nu_s|-s\nu_s\bigr),
\qquad
\nu_s=l+\frac12(1-s).
\label{eq:HNR-flat-spectrum}
\end{equation}
Thus, the relativistic flat-space spectrum used to validate the finite-element
calculation reduces to the expected planar oscillator spectrum, while the
additional terms in Eqs.~\eqref{eq:HNR-scalar}--\eqref{eq:HNR-z} identify the
leading low-energy signatures of the helical twist and of the axial torsion.

\section{Conclusions}
\label{sec:conclusions}

We have presented a complete and self-contained treatment of the Dirac oscillator
in a helically twisted spacetime carrying a uniform axial torsion. Starting from
an orthonormal vierbein and Cartan's structure equations, we obtained the
torsion-free Levi--Civita spin connection and gave the associated spinorial
matrices $\Omega_\mu^{(\mathrm{LC})}$ explicitly, both as $4\times4$ matrices and
in $2\times2$ Pauli-block form. The geometric and torsional sectors were cleanly
disentangled into two spinorial shift operators, $\hat{\mathcal S}_{\mathrm{LC}}(r)$
from the Levi--Civita connection and the constant $\hat{\mathcal S}_{\mathrm{ax}}$
from the purely axial contortion, which acts in the off-diagonal blocks of the first-order Dirac system.

Implementing the Moshinsky coupling along the local radial direction with the
matrix $\beta$, we showed that the second-order equation is not the square of a
single operator but the ordered product $\hat\Pi_+\hat\Pi_-$, and we derived the
exact radial matrix equation~\eqref{eq:radial-matrix-final-corrected}. A central
technical result is that the $1/(2r)$ contribution carried by the spin connection
combines with the oscillator term so that the spurious $-3\hbar^2/(4r^2)$ piece
cancels, leaving the correct cylindrical radial operator with the
$+\hbar^2/(4r^2)$ centrifugal term [Eq.~\eqref{eq:KpKm-final-corrected}].
The helical geometry leaves three distinct imprints on the dynamics: the
replacement of the angular momentum by the combination $m/r-\omega k$, a
geometry-induced Coulomb-like term $\propto\omega km/r$ originating from the
off-diagonal metric component through the anticommutator
$\{\gamma^\phi,\gamma^z\}=+2\omega/r\,\mathbb{I}_4$, and a $\sigma_y$ coupling
$\propto k$ that forbids a direct scalar reduction.

Substituting the separated solution, the radial problem takes the form of a
coupled, self-adjoint system of two second-order
equations~\eqref{eq:coupled-system} for the spinor components $\psi_{A,1}$ and
$\psi_{A,2}$, which we solved with a finite-element scheme that respects the
radial measure $r\,dr$. The method was validated against the analytically known
planar Dirac-oscillator spectrum~\eqref{eq:analytic-flat}, recovered exactly in
the flat limit ($\omega\to0$, $S^0\to0$, $k=0$) to a relative accuracy of order
$10^{-5}$. We computed the bound-state energies, the radial probability densities
(Fig.~\ref{fig:density}), and the dependence of the spectrum on the helical twist
$\omega$, the longitudinal momentum $k$, the oscillator frequency $\omega_D$, and
the axial torsion $\lambda_{\mathrm{ax}}S^0$ (Fig.~\ref{fig:spectra}). The
spectrum is markedly asymmetric under $k\to-k$ and displays avoided crossings,
the direct fingerprint of the angular--longitudinal mixing enforced by the
geometry, while the lowest level stays close to $E=Mc^2$, the relativistic remnant of the supersymmetric structure of the Dirac oscillator, whose fate under the deformations is analyzed below.

A further outcome of the explicit solution concerns the algebraic structure of the
model. The Dirac-oscillator supersymmetry, which pins a zero mode at $E=Mc^2$, is
preserved by the axial torsion and by the longitudinal momentum, but is softly broken
by the helical twist, with the ground-state energy departing from the rest energy as
$\Lambda_0\simeq\omega_D\omega^2/16$ (Sec.~\ref{sec:susy}). The associated
Witten-index analysis (Sec.~\ref{sec:index}) shows that the zero mode is an
unpaired supersymmetric state in the flat radial sector, and that only the helical
$1/r$ spin-connection profile removes the protected kernel of $\hat\Pi_-$. Using the
computed spectrum, we also obtained sector-resolved thermodynamic functions,
including free energy, internal energy, entropy, and heat capacity, whose
heat-capacity peak and parameter dependence (Sec.~\ref{sec:thermo}) provide
concrete predictions for fixed angular--longitudinal sectors of relativistic
fermions confined in a twisted background. The explicit spinors further determine
longitudinal vector and axial currents (Sec.~\ref{sec:currents}): the exact zero
mode is axially polarized but carries no vector current, whereas the helical
$k$-asymmetry and the axial torsion generate a finite sector-resolved current
responses. The nonrelativistic expansion (Sec.~\ref{sec:nrlimit}) further shows
that the helical twist and the axial torsion survive at low energies through an
emergent Coulomb-like potential, a renormalized centrifugal barrier, and explicit
spin-dependent geometric couplings. Several extensions follow naturally from
this framework. More general torsion configurations, including spatial components of the pseudovector $S^\mu$, would break part of the cylindrical symmetry and could be analyzed perturbatively. The inclusion of external magnetic fields or an
Aharonov-Bohm flux along the axis, the study of scattering and continuum states,
and many-sector thermodynamic extensions are all accessible within the same
formalism. Finally, because helical and torsional backgrounds can be engineered in topological materials and metamaterials, the coupled spinorial structure reported here may offer concrete signatures, most notably the $k$-asymmetry of the spectrum, for condensed-matter analogs of relativistic fermions in twisted geometries.

\section*{Acknowledgments}
M.D.M. thanks Gabriel O. Cavaleiro for fruitful discussions.
This work was partially supported by the Brazilian agencies Conselho Nacional de Desenvolvimento Cient\'{i}fico e Tecnol\'{o}gico (CNPq), Funda\c c\~ao de Amparo \`a Pesquisa e ao Desenvolvimento Cient\'{i}fico e Tecnol\'{o}gico do Maranh\~ao (FAPEMA), and Coordena\c c\~ao de Aperfei\c coamento de Pessoal de N\'{i}vel Superior (CAPES). E.O.S. acknowledges the support from CNPq (grants 306308/2022-3), FAPEMA (grant UNIVERSAL-06395/22), and CAPES (Finance Code 001).
F.M.A. acknowledges financial support from Fundação Araucária Project No. 305 and CNPq Grant No. 313124/2023-0.\\

\noindent\textbf{Conflict of interest}\\
The authors declare no conflicts of interest.\\

\noindent\textbf{Data availability}\\
No datasets were generated or analyzed during the current study.

\bibliographystyle{apsrev4-2}
%\bibliography{References}
%apsrev4-2.bst 2019-01-14 (MD) hand-edited version of apsrev4-1.bst
%Control: key (0)
%Control: author (72) initials jnrlst
%Control: editor formatted (1) identically to author
%Control: production of article title (-1) disabled
%Control: page (0) single
%Control: year (1) truncated
%Control: production of eprint (0) enabled
%

\end{document}